\documentclass[aps,prb,twocolumn,amsmath,amssymb,nofootinbib,superscriptaddress,floatfix]{revtex4}

\usepackage{bm} 
\usepackage{graphicx}
\usepackage{amssymb}
\usepackage{curves}
\usepackage{amsmath}
\usepackage{fancybox}
\usepackage{dcolumn} 
\usepackage[dvips]{color}
\usepackage{epic}
\usepackage{longtable}

\def\etal{~\textit{et~al.}} 

\def\Hc{{\rm h.c.}}

\begin{document}

\title{
Algebraic vortex liquid theory of a quantum antiferromagnet on the kagome lattice }

\author{S.\ Ryu}
\affiliation{Kavli Institute for Theoretical Physics,
	     University of California, 
	     Santa Barbara, 
	     CA 93106, 
	     USA}
\author{O.\ I.\ Motrunich}
\affiliation{Department of Physics,
             California Institute of Technology,
             Pasadena,
	     CA 91125,
	     USA}

\author{J.\ Alicea}
\affiliation{Department of Physics,
             University of California, 
             Santa Barbara, 
             CA 93106,
	     USA}

\author{Matthew\ P.\ A.\ Fisher}
\affiliation{Kavli Institute for Theoretical Physics,
	     University of California, 
	     Santa Barbara, 
	     CA 93106, 
	     USA}

\date{\today}

\begin{abstract}
There is growing evidence from both experiment and
numerical studies that low half-odd integer quantum spins
on a kagome lattice with predominant antiferromagnetic near neighbor interactions
do not order magnetically or break lattice symmetries even at temperatures much lower than the exchange interaction strength.  Moreover, there appear to be a plethora of
low energy excitations, predominantly singlets but also spin carrying,
which suggest that the putative underlying quantum spin liquid
is a gapless ``critical spin liquid'' rather than a gapped spin liquid with
topological order.  Here, we develop an effective field theory approach
for the  spin-1/2 Heisenberg model with
easy-plane anisotropy on the kagome lattice.
By employing a vortex duality transformation, followed by a fermionization and flux-smearing, we obtain access to a gapless yet stable critical spin liquid phase,
which is described by
(2+1)-dimensional quantum electrodynamics (QED$_3$) with
an emergent $\mathrm{SU}(8)$ flavor symmetry.
The specific heat, thermal conductivity, and
dynamical structure factor are extracted from the effective field
theory, and contrasted with other theoretical approaches to the kagome antiferromagnet.
\end{abstract}

\maketitle

\section{introduction} 
Kagome antiferromagnets are among the most extreme examples of
frustrated spin systems realized with nearest-neighbor interactions. 
Both the frustration in each triangular unit and the rather loose
``corner-sharing'' aggregation of these units into the kagome lattice
suppress the tendency to magnetically order. 
On the classical level, kagome spin systems are known to exhibit rather special properties: nearest-neighbor Ising and XY models
remain disordered even in the zero-temperature limit, while the $\mathrm{O}(3)$ system undergoes order-by-disorder into a coplanar spin structure.\cite{Huse92}

Quantum kagome antiferromagnets, which are much less understood, provide a fascinating arena for the
possible realization of spin liquids.
Early exact diagonalization and 
series expansion studies\cite{Zeng90, Singh92, Leung93, Elstner94}
as well as further exhaustive numerical works,
\cite{Lecheminant97, Waldtmann98, Sindzingre00, Misguich05}
provide strong evidence
for the absence of any magnetic order or other symmetry breaking in the nearest neighbor spin-1/2 system.   Moreover, a  plethora of low
energy singlet excitations is found, below a small (if non-zero) spin gap.
But the precise nature of the putative spin liquid phase in this model
has remained elusive.

On the experimental front, several quasi-two dimensional materials with magnetic moments in a kagome arrangement
have been studied, including SrCr${ }_{8-x}$Ga${}_{4+x}$O$_{19}$ with Cr${}^{3+}$ ($S=3/2$) moments\cite{Obradors88,Ramirez90,Broholm90,Ramirez92,Ramirez00},
jarosites KM${}_3$(OH)${}_6$(SO$_4$)$_2$
with M$=$Cr${}^{3+}$ or Fe${}^{3+}$
($S=5/2$) moments\cite{Wills01,Inami00,Lee97},
and volborthite Cu${}_3$V${}_2$O${}_7$(OH)${}_2\cdot$ 2H${}_2$O with
Cu${}^{2+}$ ($S=1/2$) moments\cite{Hiroi01}.
Recently,
herbertsmithite ZnCu${}_3$(OH)${}_6$Cl${}_2$ which also
has Cu${}^{2+}$  moments on a kagome lattice
was synthesized for which
no magnetic order is observed down to 50 mK 
despite the estimated exchange constant of 300~K.
\cite{Helton06,Ofer06,Mendels06}
The second layer of $^{3}$He absorbed on graphoile
is also believed to realize the kagome magnet.\cite{Masutomi04, Elser89}
The suppression of long-range spin correlations or other signs of symmetry breaking down to temperatures
much lower than the characteristic exchange energy scale is manifest in
all of these materials, consistent with expectations.
Moreover, there is evidence for low energy excitations, both spin carrying and singlets.
But the ultimate zero-temperature spin liquid state is often masked in
these systems by magnetic ordering or glassy behavior at the lowest temperatures, perhaps due
to additional interactions or impurities, rendering the experimental study of the spin liquid properties problematic.
New materials and other experimental developments are changing this situation, and the question of the quantum spin liquid ground state
of the kagome antiferromagnet is becoming more prominent.

There are two broad classes of spin liquids which have been explored theoretically,
both in general terms and for the kagome antiferromagnet in particular.
The first class comprise the ``topological'' spin liquids, 
which have a gap 
to all excitations and have particle-like excitations with fractional quantum numbers above the
gap.   Arguably, the simplest topological liquids are the so-called $Z_2$ spin liquids, 
which support a vortex like excitation - a vison - in addition 
to the spin one-half spinon.  For the kagome antiferromagnet,
Sachdev~\cite{Sachdev91} and more recently
Wang and Vishwanath~\cite{Wang06} 
have employed a Schwinger boson approach to systematically
access several different $Z_2$ spin liquids.  
For a kagome antiferromagnet with easy axis anisotropy and further
neighbor interactions, Balents\etal\cite{Balents02}
unambiguously established the presence
of a $Z_2$ spin liquid, obtaining an exact ground state wavefunction
in a particular limit.    
Quantum dimer models on the kagome lattice can also support
a $Z_2$ topological phase\cite{Misguich02}.
Spin liquids with topological order and time reversal
symmetry breaking, the chiral spin liquids which are closely analogous to fractional quantum Hall states, have been found on the kagome lattice\cite{Marston91,KunYang93}
within a  fermionic representation of the spins.
But all of these topological liquids are gapped, and cannot account for
the presence of many low excitations found in the exact diagonalization studies
and suggested by the experiments.

 A second class of spin liquids - the ``critical'' or ``algebraic spin liquids'' (ASL) - have gapless
 singlet and spin carrying excitations.  Although these spin liquids share
 many properties with  quantum critical points such as correlation
 functions falling off
as power laws with non-trivial exponents, they are believed to be stable
quantum phases of matter.  
Within a fermionic representation of the spins,  Hastings\cite{Hastings01} explored
an algebraic spin liquid on the kagome lattice, and more
recently Ran\etal\cite{Ran06} have extended his analysis 
in an attempt to explain the observed properties of herbertsmithite ZnCu$_3$(OH)$_6$Cl$_2$.

In this paper, motivated by the experiments on the kagome materials and the numerical studies, we pursue yet a different approach to the possible
spin liquid state of the kagome antiferromagnet.
As detailed below, for a kagome antiferromagnet with easy-plane anisotropy we find evidence for
a new critical spin liquid, an    
``algebraic vortex liquid'' (AVL) phase.  Earlier we had
introduced and explored the AVL in the context of the triangular XXZ
antiferromagnet in 
Refs.\ \onlinecite{Alicea05,Alicea05a,Alicea05b}.
Compared to the slave particle techniques for studying  frustrated
quantum antiferromagnets,  the AVL approach is less microscopically
faithful to a specific spin Hamiltonian, but has the virtue of being unbiased.

Our approach requires the presence
of an easy-plane anisotropy, which for a spin one-half system can
come from the Dzyaloshinskii-Moriya interaction, although this is usually quite small.
But for higher 
half-integer spin systems, spin-orbit coupling allows for a single
ion anisotropy $D (S^z)^2$ which can be appreciable.  With $D>0$ 
this leads to an easy-plane spin character, and from a symmetry stand point our analysis
should be relevant.   Moreover, 
a very interesting unpublished exact diagonalization study by 
Sindzingre \cite{Sindzingre}
suggests that an unusual spin liquid phase is also realized  for
the nearest-neighbor quantum XY antiferromagnet.
He finds a small gap to $S^z=1$ excitations, but below this gap
there is a plethora of $S^z=0$ states, which is reminiscent of the many
singlet excitations below the triplet gap in the
Heisenberg SU(2) spin model.\cite{Lecheminant97}
Thus, based on the exact diagonalization studies,
the easy plane anisotropy does not
appear to gap out the putative critical spin liquid of the Heisenberg model,
although it will presumably modify its detailed character.

Our approach to easy-plane frustrated quantum antiferromagnets
focuses on vortex defects in the spin configurations, rather than the 
spins themselves.
In this picture, the magnetically ordered phases do not have vortices
present in the ground state.  But there will be gapped vortices,
and these can lead to important effects on the spectrum.
For example, vortex-antivortex (i.e. roton) excitations can lead to minima
in the structure factor at particular wavevectors in the 
Brillouin zone,
\cite{Zheng06,Zheng06b,Starykh} in analogy to superfluid He-4.
As we demonstrated for the triangular antiferromagnet and explore below
for the kagome lattice, our approach predicts specific locations in the 
Brillouin zone for the roton minima, which could perhaps be checked by
high order spin wave expansion techniques.
On the other hand, the presence
of mobile vortex defects in the ground state can destroy magnetic order,
giving access to various quantum paramagnets.
If the vortices themselves condense, the usual result is the breaking of
lattice symmetries, such as in a spin Peierls state.  But if the 
vortices remain gapless one can access a critical spin liquid phase.

For frustrated spin models the usual duality transformation to vortex
degrees of freedom does not resolve the geometric frustration since the vortices are
at finite density.   However, by binding $2\pi$-flux to each vortex converting them into fermions coupled to a Chern-Simons gauge field followed by a simple flux-smearing mean-field treatment gives a simple way to describe the vortices.
The long-range interaction of vortices actually works to our advantage
here since it suppresses their density fluctuations and leads
essentially to incompressibility of the vortex fluid.  
As demonstrated in the easy-plane quantum 
antiferromagnet on the
triangular lattice,\cite{Alicea05a, Alicea05b} including fluctuations about 
the flux-smeared mean field enables one to access
a critical spin liquid with gapless vortices.  The theory has the structure
of a (2+1)-dimensional [(2+1)D]
quantum electrodynamics (QED$_3$), with relativistic fermionic vortices minimally
coupled to a non-compact U(1) gauge field.
The resulting  algebraic vortex liquid (AVL) phase
is a novel critical spin liquid phase that exhibits neither magnetic nor any other symmetry-breaking order.  This approach also allows one to study many competing orders in the vicinity of the gapless phase.

When applied to the spin-1/2 easy-plane antiferromagnet
on the kagome lattice,
the duality transformation combined with
fermionization and flux smearing also leads to
a low-energy effective QED$_3$ theory with 8 flavors of Dirac fermions with an
emergent $\mathrm{SU}(8)$ flavor symmetry.
Amusingly, AVL's with
$\mathrm{SU}(2)$ \cite{Alicea05},
$\mathrm{SU}(4)$ \cite{Alicea05a, Alicea05b} and
$\mathrm{SU}(6)$ \cite{Alicea06}
emergent
symmetries were obtained previously
for quantum XY antiferromagnets on the triangular lattice,
with integer spin, half-integer spin and half-integer spin in an applied magnetic field,
respectively.
QED$_{3}$ theory with $N$ flavors of
Dirac fermions is known to realize a stable critical phase for
sufficiently large $N > N_c$. 
While numerical attempts to determine $N_c$ are so far inconclusive,\cite{Hands02, Hands04} 
an estimate from the large-$N$ expansion suggests 
$N_c \sim 4$.\cite{Alicea05} 
It seems very likely that  $N=8$ is large enough, implying the presence of a stable 
critical spin liquid ground state for the easy-plane
spin-1/2 quantum antiferromagnet on the
kagome lattice.

Although the algebraic vortex liquid and the algebraic spin liquid\cite{Hastings01,Ran06} obtained for the kagome lattice are accessed in rather different
ways, they share a number of commonalities, both theoretically and with regard
to their experimental implications.  Both approaches end with a QED$_3$
theory, the former a non-compact gauge field theory with fermionic vortices carrying an emergent SU(8) symmetry, and the
latter a compact gauge theory with fermionic spinons with SU(4) symmetry.
The non-compact nature of the gauge field in the AVL follows from 
the fact that the $S^z$ component of spin appears as the gauge flux
in the dualized theory. 
Hence the states with zero flux and $2\pi$ flux
are physically distinct; moreover, it
follows that since the total $S^z$ is conserved, so also is the total gauge
flux. This should be contrasted with the slave particle approach, where a
compact gauge theory arises on the lattice and there is no such
conservation law. Physically, this means that dynamical monopole
operators, which could potentially destabilize the spin liquid and open a
gap in the excitation spectrum, are not allowed in our low-energy theory
for the AVL, while they are allowed in a low-energy description of spin
liquids obtained using a slave particle framework.

With regard to experimentally accessible quantities, both theories predict a power law specific heat $C \sim T^2$ at low temperatures,
which follows from the linear dispersion of the
fermions, and which would dominate
the phonon contribution to the specific heat for magnets
with exchange interactions significantly smaller that the Debye frequency.
\cite{remark specific heat}
Because the spinons  couple directly to magnetic fields,
one would expect the specific heat in the ASL to be more sensitive to an external applied
magnetic field than the AVL phase.
Both theories predict a thermal conductivity which vanishes as 
$\kappa \sim T$,
\cite{Ioffe90, Lee92} which is an interesting experimental signature
reflecting the dynamical mobility of the gapless excitations.

The momentum resolved dynamical spin structure factor
which can be extracted from 
inelastic neutron experiments can in principle give very detailed information
about the spin dynamics.  
Following the framework developed on the
triangular lattice,\cite{Alicea05, Alicea05a, Alicea05b,Alicea06}
for the kagome AVL we can extract
the wavevectors in the magnetic Brillouin  zone which have
gapless spin carrying excitations, and find 12 of them as shown
in  Fig.~\ref{fig: enhbilin.eps}.  In contrast, the kagome ASL phase is predicted to have gapless spin excitations at only 4 of these 12 momenta.
The momentum space location of the gapless excitations is 
perhaps the best way to try and distinguish experimentally between
these two (and any other) critical  spin liquids.

The rest of the paper is organized as follows.
We start by introducing our model
in Sec.\ \ref{sec: model}.
The low-energy effective field theory 
is then developed in
Sec.\ \ref{sec : Fermionized vortex description}.
In Sec.\ \ref{sec : Properties of the AVL phase},
the properties of the critical spin liquid phase (AVL phase)
are discussed,
especially,
the spin excitation (roton spectrum) (Fig.\ \ref{fig:spinwaves}),
the $S^z$ dynamical structure factor (Eq.~\ref{eq: Sz structure factor}),
and the in-plane dynamical structure factor (Eq.~\ref{eq : S^+- structure factor}).
We conclude in
Sec.\ \ref{sec: conclusion}.
Some details of the analysis are presented in three 
Appendices. 

\section{Extended kagome Model}
\label{sec: model}

Our primary interest is the kagome lattice spin-1/2 model with
easy-plane anisotropy (quantum XY model)
\begin{eqnarray}
\mathcal{H}
&=&
\frac{1}{2}
\sum_{\mathbf{r},\mathbf{r}'}
\left[
J^{\ }_{\mathbf{r},\mathbf{r}'}
S^+_{\mathbf{r}}
S^-_{\mathbf{r}'}
+
\mathrm{h.c.}
\right]
+
\sum_{\mathbf{r},\mathbf{r}'}
J^{z}_{\mathbf{r},\mathbf{r}'}
S^z_{\mathbf{r}}
S^z_{\mathbf{r}'},
\end{eqnarray}
with $J^{\ }_{\mathbf{r},\mathbf{r}'} > J^z_{\mathbf{r},\mathbf{r}'}$. 
We consider the system with dominant
nearest-neighbor exchange $J_1$, but will also allow some second- and
third-neighbor exchanges $J_2$ and $J_3$.

\begin{figure}
  \begin{center}
  \includegraphics[width=6cm,clip]{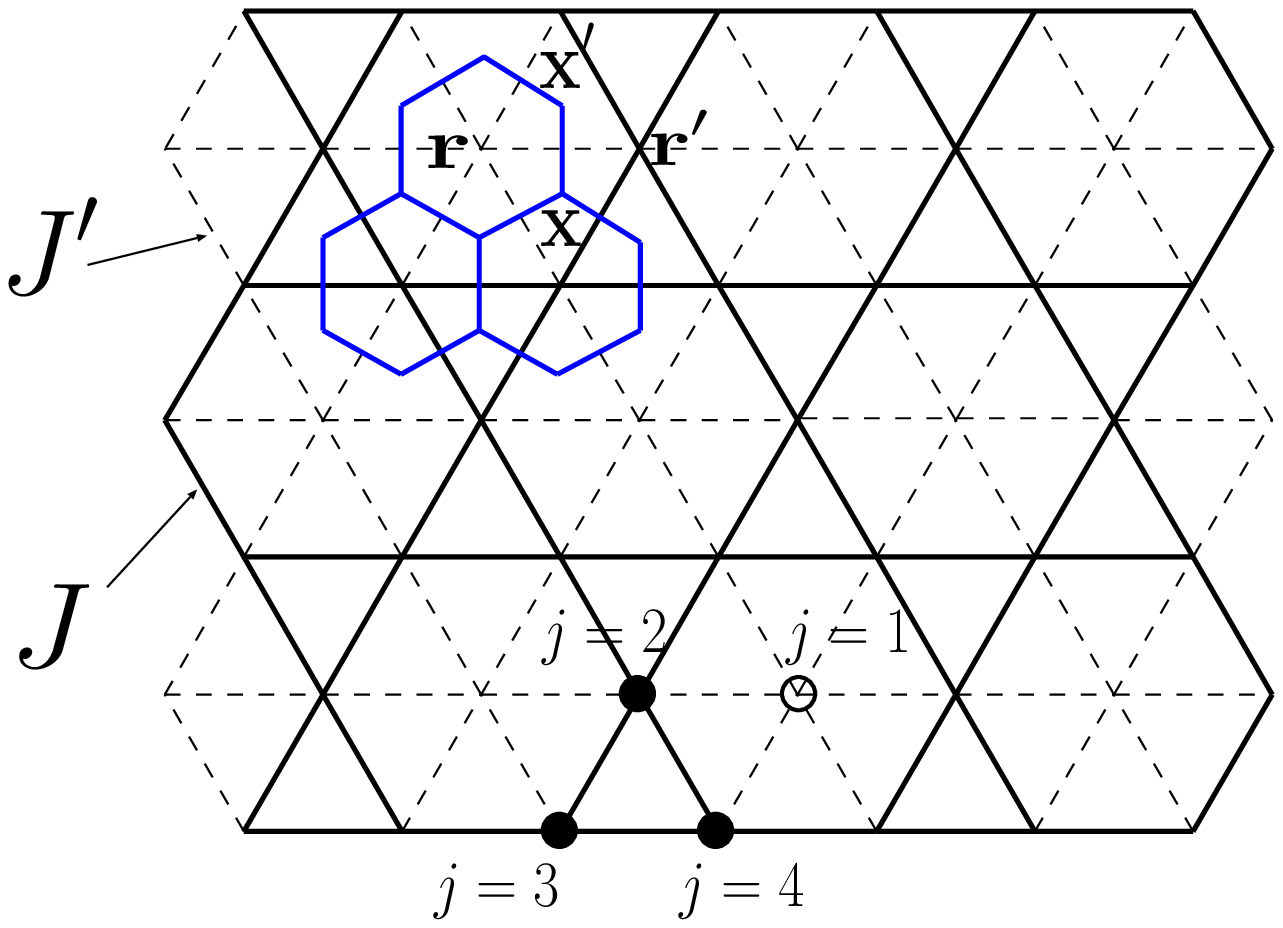} \\
\vspace{0.5cm}
  \includegraphics[width=3.8cm,clip]{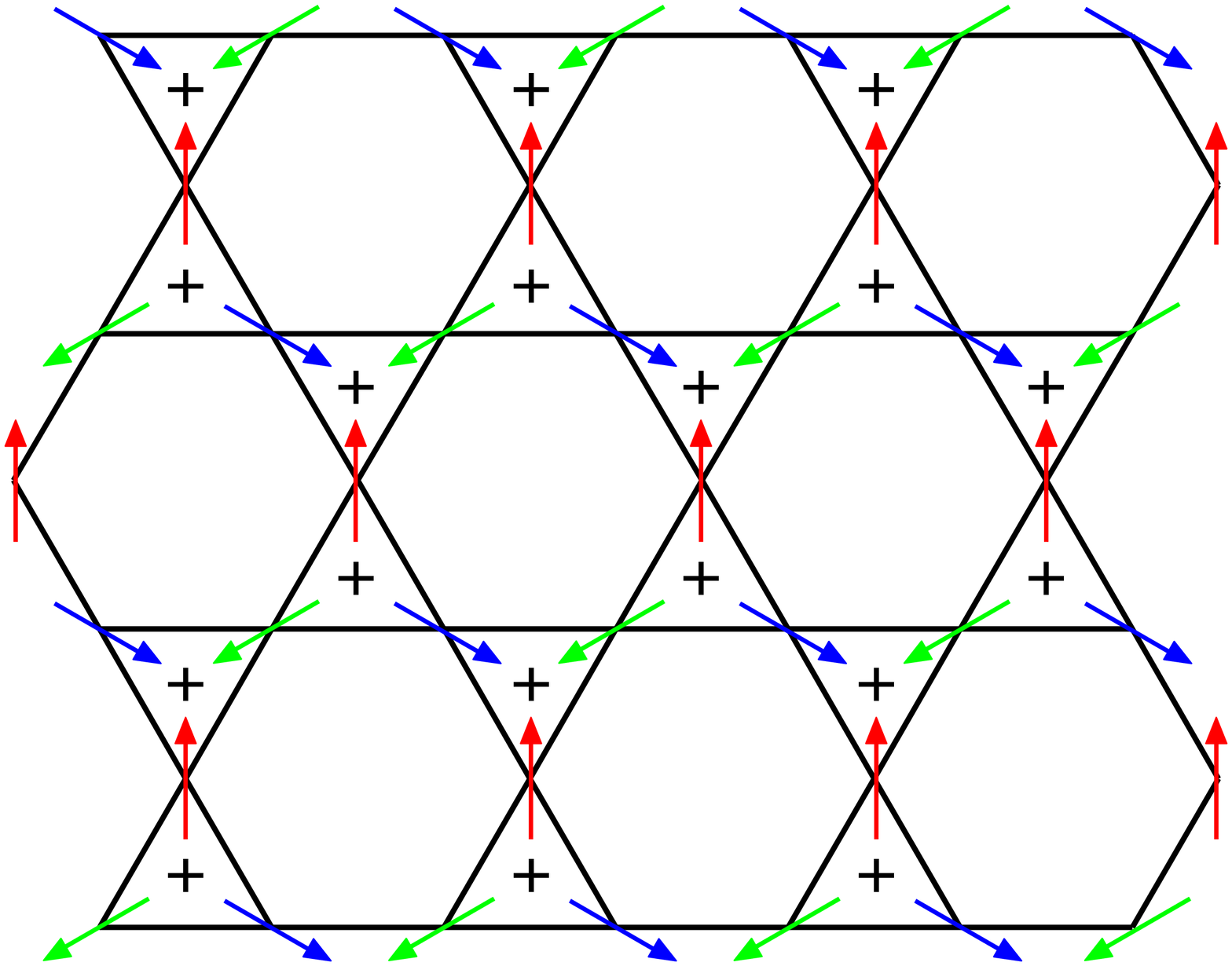}
  \includegraphics[width=3.8cm,clip]{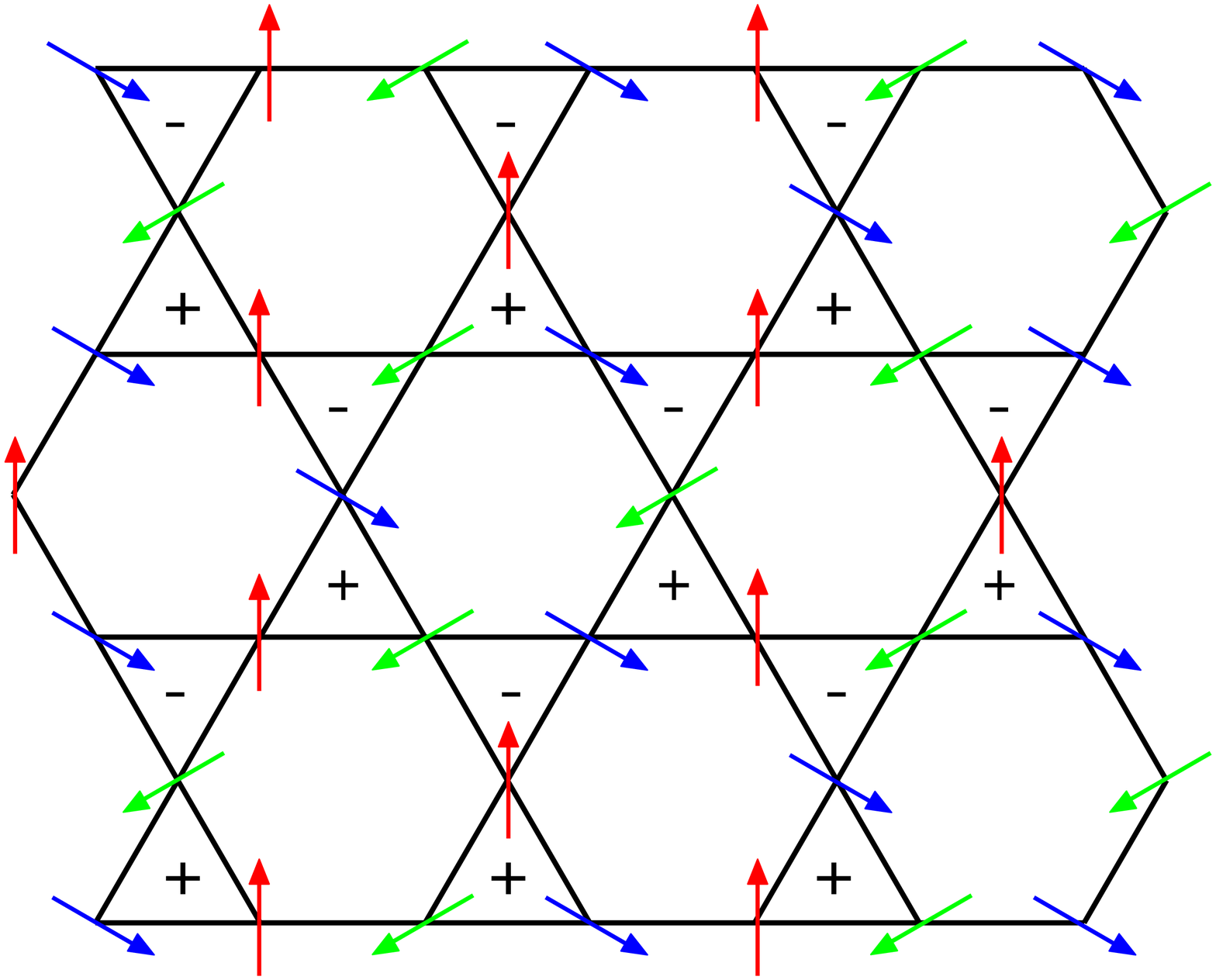}
\caption{
\label{fig: kagome_lattice}
(Top) The quantum spin model with easy-plain anisotropy,
Eq.\ (\ref{eq: definition of the model}),
on the kagome lattice  (thick line) supplemented
by an extra site at the center of each hexagon.
There is a half-integer spin for each kagome site
whereas integer spins are placed at each center of hexagons.
Four spins in a unit cell are labeled by $j=1,2,3,4$.
For this triangular extension of the kagome lattice, 
the dual lattice is the honeycomb lattice indicated in the upper left part.
(Bottom)
The $q=0$ state (Left) and the $\sqrt{3}\times \sqrt{3}$ state (Right).
Positive/negative vorticities (chiralities) for each triangle are denoted by
$+/-$.
}
\end{center}
\end{figure}

To set the stage, the model with only nearest-neighbor coupling has
an extensive degeneracy of classical ground states, which is lifted
by further-neighbor interactions.  For example, with antiferromagnetic
$J_2 > 0$, the so-called $q=0$ phase is stabilized
(Fig.~\ref{fig: kagome_lattice}).
On the other hand,
for ferromagnetic $J_2 < 0$, the so-called $\sqrt{3}\times \sqrt{3}$
structure is the classical ground state
(Fig.~\ref{fig: kagome_lattice}).
These ordered phases are also realized in the quantum spin-1/2 
Heisenberg
model for large enough $J_2$\cite{Lecheminant97},
while the
case with $J_2/J_1 \simeq 0$ is most challenging as mentioned in the
introduction.

The easy-plane spin system can be readily reformulated in terms of
vortices.  But in order to apply the fermionized vortex approach
more simply, we consider a wider class of models which
includes the nearest-neighbor kagome model.
Specifically, we add one extra site at the center of each hexagon,
on which we put an integer spin
(Fig.~\ref{fig: kagome_lattice}).
(Technical reasons for doing this are discussed at the end of 
Sec.~\ref{sec : Fermionized vortex description}.)
In the rotor representation of spins, with a phase $\varphi$ canonically conjugate to the integer boson number, $n \sim S^z + 1/2$ on the kagome sites and
$n \sim S^z$ at the centers of each hexagon, such an extended model reads
\begin{eqnarray}
  \mathcal{H}
  &=&
  \sum_{\mathbf{r},\mathbf{r}'}J_{\mathbf{r},\mathbf{r}'}
  \cos\left(
    \varphi_{\mathbf{r}}
    -
    \varphi_{\mathbf{r}'}
  \right)
  +
  \sum_{\mathbf{r}} U_{\mathbf{r}}
  \left(
    n^{\ }_{\mathbf{r}}
    -n^0_{\mathbf{r}}
  \right)^2,
\label{eq: definition of the model}
\end{eqnarray}
where $n^0 = 1/2$ for the kagome lattice sites,
while $n^0 = 0$ for the added hexagon center sites.
We have dropped the $J^z$ term for simplicity 
since it amounts to a mere renormalization of the interactions
in the effective field theory that we will derive.
As shown in Fig.\ \ref{fig: kagome_lattice}, we take the interaction
to be $J$ for the nearest neighbor half-integer spins,
while the coupling that connects integer and half-integer spins is $J'$.
When the on-site interaction $U$ is infinitely large, $U\to \infty$,
there remain two low-energy states realizing the Hilbert space of a
$S=1/2$ spin at the kagome sites,
whereas the frozen sector with $n_{\mathbf{r}}=0$
is selected at the hexagon center sites.
In the quantum rotor model, we ``soften'' this constraint and take $U$
to be finite.

The additional integer-spin degrees of freedom do not spoil 
any symmetries of the original model, which for the record include lattice translations, reflections, and rotations, as well as XY spin symmetries.  
If we integrate over the extra degrees of
freedom, the model looks very close to the original one.
For example, in the limit when $J' \ll U$, we obtain
the kagome model with additional ferromagnetic second and third neighbor
exchanges $J_2 = J_3 = - (J')^2/(2U)$,
together with a renormalized nearest-neighbor exchange
$J_1 = J - (J')^2/(2U)$.
Such exchanges would slightly favor the $\sqrt{3}\times\sqrt{3}$ state
in the classical limit, but as we will see, the spin liquid phase
that we obtain by analyzing the extended kagome model will have
enhanced correlations corresponding to both the $q=0$ and
$\sqrt{3}\times\sqrt{3}$ phases that are nearby.  Thus the small bias introduced by the added integer-spin sites appears to be not so important for the generic spin liquid phase that we want to describe.

\section{Fermionized vortex description}
\label{sec : Fermionized vortex description}

\begin{figure}
  \begin{center}
  \includegraphics[width=7cm,clip]{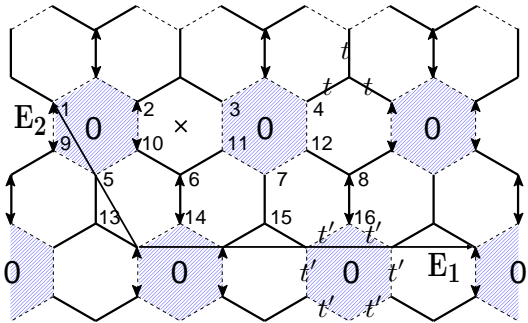}
\caption{
\label{fig: honeycomb_lattice_flux_lesik}
The flux-smeared mean field background for fermionized vortices.  Each hexagon is
threaded by either zero or $\pi$-flux.  (hexagons with zero-flux are
specified by ``0'' whereas all the other hexagons are pierced by
$\pi$-flux.)  A convenient choice of a gauge is
also shown: for links labeled by $\leftrightarrow$, we assign the Peierls
phase factor $e^{-{i}a^0_{\mathbf{x}\mathbf{x}'}}=e^{{i}\pi}$.  The unit cell in this gauge
consists of 16 sites as labeled in the figure.  
For links represented by a solid line 
we assign hopping amplitude $t(=1)$ 
whereas 
for links denoted by a broken line we assign hopping $t'$ .
We take $\times$ in the figure as an origin.
}
\end{center}
\end{figure}

The triangular extension of the kagome XY antiferromagnet
discussed above is somewhat similar to the triangular lattice spin-1/2 XY antiferromagnet
studied in Refs.\ \onlinecite{Alicea05a, Alicea05b}, except that
one quarter of the sites is occupied by integer spins. 
The duality transformation for the present model proceeds
identically to Refs.\ \onlinecite{Alicea05a, Alicea05b},
and turns the original spin Hamiltonian
(\ref{eq: definition of the model}) into
the dual Hamiltonian describing vortices
hopping on the honeycomb lattice and interacting
with a gauge field residing on the dual lattice links,
\begin{eqnarray}
\mathcal{H} &=&
2\pi^2
\sum_{\mathbf{x}\mathbf{x'}}
J_{\mathbf{x}\mathbf{x'}}
e^{2}_{\mathbf{x}\mathbf{x'}}
+
\frac{U}{(2\pi)^2}
\sum_{\mathbf{r}}
\left(
\nabla \times a
\right)_{\mathbf{r}}^{2}
\nonumber \\
&&
-2
\sum_{\mathbf{x}\mathbf{x'}}
t_{\mathbf{x}\mathbf{x'}}
\cos
\left(
\theta_{\mathbf{x}}
-
\theta_{\mathbf{x'}}
-
a^{\ }_{\mathbf{x}\mathbf{x}'}
-
a^{0}_{\mathbf{x}\mathbf{x}'}
\right).
\label{eq: final dual Hamiltonian for bosonic vortices}
\end{eqnarray}
Here $\mathbf{x},\mathbf{x}'$ label sites
on the honeycomb lattice (see Fig.~\ref{fig: kagome_lattice}),
$e^{\pm i \theta_{\mathbf{x}}}$
is a vortex creation/annihilation operator at site $\mathbf{x}$,
and
$a_{\mathbf{x}\mathbf{x}'},
 e_{\mathbf{x}\mathbf{x}'}$
represent a gauge field on the link
connecting $\mathbf{x}$ and $\mathbf{x}'$.
$J_{\mathbf{x}\mathbf{x}'}$ is given by
$J_{\mathbf{x}\mathbf{x}'}=J (J')$ when 
the dual link $\langle \mathbf{x}\mathbf{x}' \rangle$
crosses the original lattice link $\langle \mathbf{r}\mathbf{r}' \rangle$
with $J_{\mathbf{r}\mathbf{r}'}=J (J')$.
The last term in the dual Hamiltonian represents
the vortex hopping where
the hopping amplitude 
$t_{\mathbf{x}\mathbf{x}'}$ is given by
$t_{\mathbf{x}\mathbf{x}'}=t (t')$ when 
the link $\langle \mathbf{x}\mathbf{x}' \rangle$
crosses the link $\langle \mathbf{r}\mathbf{r}' \rangle$
with $J_{\mathbf{r}\mathbf{r}'}=J (J')$.
Crudely, we have $t/t' \sim J'/J$ since vortices hop more easily
across weak links (Fig.\ \ref{fig: honeycomb_lattice_flux_lesik}).
Finally, the static gauge field $a^0$ encodes the average flux
seen by the vortices when they move; 
this is described in more detail below.

In the dual description,
because of the frustration in the spin model,
the average density of vortices 
is
one half per site. 
On the other hand, 
the original boson density
is viewed as
a gauge flux
\begin{eqnarray}
&&
  n_{\mathbf{r}}
=
  \frac{1}{2\pi}\left(
    \nabla \times \boldsymbol{a}
  \right)_{\mathbf{r}}
=
  \frac{1}{2\pi}\sum_{\begin{tiny}
\mbox{
$\langle \mathbf{x}\mathbf{x}'\rangle$ around $\mathbf{r}$}
\end{tiny} }
a_{\mathbf{x}\mathbf{x}'} .
\label{eq: dual transformation}
\end{eqnarray}
Therefore, vortices experience on average $\pi$
flux going around the triangular lattice sites with half-integer spins,
but they see zero flux going around the sites with integer spins.
Thus, one quarter of the hexagons will have zero flux as seen by the
vortices, and this is where the present model departs from the
considerations in Refs.\ \onlinecite{Alicea05a, Alicea05b}.

To treat the system of interacting vortices at finite density,
we focus on the two low-energy states 
with vortex number
$N_{\bf x} = 0$ and $N_{\bf x} = 1$ at each site and view vortices as 
hard-core bosons.  We can then employ the fermionization as
in Refs.~\onlinecite{Alicea05a,Alicea05b}
and arrive at the following hopping Hamiltonian for fermionized vortices 
$d_{\bf x}$
\begin{eqnarray}
\mathcal{H}_{\mathrm{ferm}} 
=
-
\sum_{\mathbf{x}\mathbf{x'}}
\left[
 t_{\mathbf{x}\mathbf{x'}}
d^{\dag}_{\mathbf{x}}
d^{\ }_{\mathbf{x'}}
e^{-{i}
\left(
a_{\mathbf{x}\mathbf{x}'}
+
a_{\mathbf{x}\mathbf{x}'}^{0}
+
A_{\mathbf{x}\mathbf{x}'}
\right)}
+
\mathrm{h.c.}
\right],
\end{eqnarray}
where we have introduced a Chern-Simons field $A$ whose flux is tied to 
the vortex density, $(\nabla \times A)_{\mathbf{x}} = 2\pi N_{\mathbf{x}}$.

Before proceeding with the analysis of the fermionized vortex
Hamiltonian, we now point out the technical reasons for considering
the extended kagome model.
If we apply the duality transformation to the kagome model with
nearest-neighbor antiferromagnetic coupling $J_1$ only,
we would obtain a dual vortex theory on the dice lattice,
which is the dual of the kagome lattice.
In this theory, in order to reproduce the rich physics while
restricting the vortex Hilbert space at each site to something
more manageable like the hard-core vortices described earlier,
we would need to keep two low-energy states of vortices for each
triangle of the kagome lattice
(i.e., for each three-fold coordinated site of the dice lattice),
whereas there are {\em three} such states to keep for each hexagon
(i.e., six-fold coordinated dice site).
The latter degrees of freedom are
rather difficult to represent in terms of
fermions. 
On the other hand, in the dual treatment of the extended model,
such a six-fold coordinated dice lattice site is effectively split into
six sites.  Each such new site now has two low-energy states but the 
sites are coupled together, and this provides some caricature of the 
original important vortex states on the problematic dice sites. 
This representation now admits standard fermionization,
and it was ``found'' in a rather natural way without any prior bias
regarding how to treat the difficult dice sites with three important vortex states.

\subsection{Flux smearing mean field and low energy effective
theory}
\label{sec: low-energy effective field theory}

The flux attachment implemented above is formally exact,
and there have been no approximations so far,
although
we have softened the discreteness constraints on the 
gauge fields.
To arrive at a low-energy effective field theory,
we now pick a saddle point configuration of the
gauge field, incorporating a flux-smeared mean-field ansatz, and later will include fluctuations
around it. 
That is, we first distribute (or ``smear'') the $2\pi$ flux attached to each
vortex on a dual site to the three (dual) plaquettes surrounding it.
Since the vortices are at half filling and each hexagon contains
six sites, the total smeared flux through each hexagon is $2\pi$,
which is equivalent to no additional flux.  
Therefore we assume $A_{\mathbf{x}\mathbf{x}'}=0$ in the mean field.
Also, we take $a_{\mathbf{x}\mathbf{x}'}=0$ on average, including the fluctuations 
after we identify the important low-energy degrees of freedom
for the fermionic vortices.
Thus, at the mean field level,
the original Hamiltonian (\ref{eq: definition of the model})
is converted into a problem of fermions hopping on the honeycomb lattice,
\begin{eqnarray}
\mathcal{H}_{\mathrm{ferm,MF}}
&=&
\sum_{\mathbf{x}\mathbf{x'}}
\left[
t_{\mathbf{x}\mathbf{x}'}
e^{-{i}
a^{0}_{\mathbf{x}\mathbf{x}'}
}
d^{\dag}_{\mathbf{x}}
d^{\ }_{\mathbf{x}'}
+
\mathrm{h.c.}
\right].
\label{eq: flux-smeared mean-field Hamiltonian}
\end{eqnarray}
In the mean
field, 3 out of 4 hexagons are pierced by $\pi$-flux, while the
remaining hexagons with an integer spin at the center
have zero flux, see Fig.\ \ref{fig:  honeycomb_lattice_flux_lesik}.
Our choice of the gauge is shown in Fig.\ \ref{fig:  honeycomb_lattice_flux_lesik}.
The unit cell consists of 16 sites,
the area of which is twice as large as the physical unit cell of the original
kagome spin system.

\begin{figure}
  \begin{center}
    \includegraphics[width=6.5cm,clip]{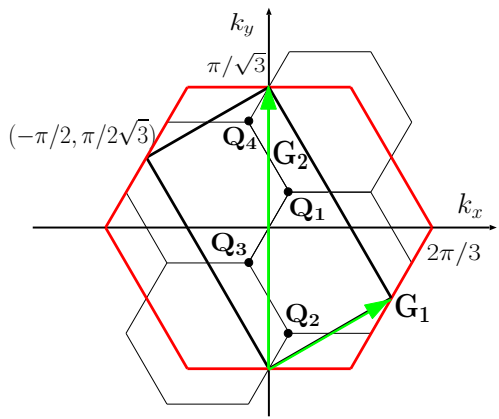}
\caption{
\label{fig: bz}
The first Brillouin zone for the flux smeared
mean field ansatz
and the location of the nodes $\mathbf{Q}_{1,\ldots, 4}$. 
The hexagonal physical Brillouin zone is also presented.
Small hexagons are guide for eyes.
}
\end{center}
\end{figure}

This Hamiltonian can be solved easily
by the Fourier transformation.
The lattice vectors that translate the unit cell are given by
\begin{eqnarray}
  \mathbf{E}_1 =
  \big(4,0 \big),
  \quad
  \mathbf{E}_2 =
  \big(-1,\sqrt{3}\big).
\end{eqnarray}
and the first Brillouin zone (BZ) is specified as
the Wigner-Seitz cell of the
reciprocal lattice vectors
\begin{eqnarray}
  \mathbf{G}_1 = \left(\frac{\pi}{2},
\frac{\pi}{2 \sqrt{3}}\right),
  \quad
  \mathbf{G}_2 = \left(0,\frac{2\pi}{\sqrt{3} }\right).
\end{eqnarray}
This is shown in Fig.\ \ref{fig: bz}.
Note that, on the other hand, the physical unit cell is defined by
$
  \mathbf{E}_1^{\mathrm{phys}}=
  \mathbf{E}_1/2 =
  \left(2,0 \right),
$
and
$
  \mathbf{E}_2^{\mathrm{phys}}=
  \mathbf{E}_2 =
  \left(-1,\sqrt{3}\right).
$
Thus the physical BZ
is determined as the Wigner-Seitz cell of
the reciprocal lattice vectors
$
  \mathbf{G}^{\mathrm{phys}}_1 = \left(\pi,\pi /\sqrt{3}\right)
$
and
$
  \mathbf{G}^{\mathrm{phys}}_2 = \left(0, 2\pi/\sqrt{3} \right)
$,
and is also shown in Fig.\ \ref{fig: bz} and 
Fig.~\ref{fig: enhbilin.eps}.

Because of the particle-hole symmetry of the original quantum rotor Hamiltonian,
which in terms of the fermionized vortices 
is realized as a vortex particle-hole symmetry 
(see Appendix \ref{app : PSG}),
the band structure of the flux-smeared mean-field Hamiltonian
(\ref{eq: flux-smeared mean-field Hamiltonian})
is symmetric around zero energy, and consists of 
eight bands with positive energy and eight with negative energy.  
The 7th and 8th bands have negative energy spectra
and are completely degenerate,\cite{Vafek} each having four Fermi points 
$\mathbf{Q}_{1,2,3,4}$ that touch
zero energy.  In the same way, the 9th and 10th bands have positive
energy spectra and are completely degenerate, and 
each of them touches zero energy at the same four Fermi points
$\mathbf{Q}_{1,2,3,4}$ thus completing the Dirac nodal spectrum.
With the gauge choice specified in Fig.\ \ref{fig:  honeycomb_lattice_flux_lesik},
the locations of the Fermi points in the BZ are
\begin{eqnarray}
\mathbf{Q}_1 = -\mathbf{Q}_3 &=& 
  \left(\frac{\pi}{12}, \frac{\pi}{4\sqrt{3}} \right), \\
\quad
\mathbf{Q}_2 = -\mathbf{Q}_4 &=&
  \left(\frac{\pi}{12}, -\frac{3\pi}{4\sqrt{3}} \right),
\label{eq: Dirac points}
\end{eqnarray}
(see Fig.\ \ref{fig: bz}).
The low-energy spectrum consists of eight gapless two-component
Dirac fermions with identical Fermi velocities given by
(here we take $t=1$ for convenience)
\begin{eqnarray}
  v_{F} =
\sqrt{
  \frac{6 t^{\prime 2}}{(3+t^{\prime 2})(3+2 t^{\prime 2})} } ~.
\end{eqnarray}
Hence
the low-energy effective theory
enjoys
an emergent $\mathrm{SU}(8)$ symmetry
among eight Dirac cones, which is protected at the kinetic-energy
level by the underlying discrete symmetries in the problem
(see Appendix~\ref{sec : Fermion bilinears}).

The lattice fermionized vortex operator is
conveniently expanded in terms of slowly varying continuum fields
$\psi^{\ }_{i\alpha a}(\mathbf{x})$ by
\begin{eqnarray}
  d_{\mathbf{x}}
  &\sim&
  \sum_{\alpha=1}^{4}
  \sum_{i=\mathrm{I},\mathrm{II}}
  \sum_{a=1,2}
  e^{{i}\mathbf{Q}_{\alpha} \cdot \mathbf{x}}
  \phi^{\ }_{i\alpha a}(n)\,
  \psi^{\ }_{i\alpha a}(\mathbf{x}),
\label{eq: def continuum field}
\end{eqnarray}
where indices
$\alpha=1, \dots, 4$,
$i=\mathrm{I},\mathrm{II}$, and
$a = 1,2$,
refer to the four Fermi momenta,
two sublattices of the honeycomb lattice, and two degenerate bands, respectively,
while $n =1,\dots, 16$ specifies the site label in the
unit cell, cf.~Fig.~\ref{fig:  honeycomb_lattice_flux_lesik}.
The 16-component wavefunctions
$  \phi^{\ }_{i\alpha a}(n)$
represent modes at the Fermi points and are specified
in Appendix~\ref{sec : Zero mode wave functions at the Fermi points}.

Working with these continuum fields, we find that the low-energy 
Hamiltonian has the same form 
$v_F (-\sigma^1 \hat p_x + \sigma^2 \hat p_y)$ near each Dirac node,
where $\sigma^{1,2,3}$ are Pauli matrices that act
on the ``Lorentz'' indices $i=\mathrm{I},\mathrm{II}$.
Correspondingly,
by introducing the gamma matrices in $(2+1)$D
\begin{eqnarray}
  &&
  \gamma_{0}=\sigma^{3},
  \quad
  \gamma_{1}=-\sigma^{2},
  \quad
  \gamma_{2}=-\sigma^{1},
\nonumber \\
&&
    \gamma_{\mu}
    \gamma_{\nu}
+
    \gamma_{\nu}
    \gamma_{\mu}
  =
  2\delta_{\mu\nu},
\quad \mu,\nu=0,1,2,
\label{eq: gamma matrices}
\end{eqnarray}
the Euclidean low-energy effective action is given by
\begin{eqnarray}
  S_{\mathrm{eff}}
  &=&
  \int d^{3}x
  \sum_{(\alpha)=1}^{8}
  \bar{\psi}^{\ }_{(\alpha)}\,
  \gamma_{\mu} \partial_{\mu}
  \psi^{\ }_{(\alpha )}
\end{eqnarray}
where $(\alpha)=\alpha a$ is a collective index running from
1 to 8;
$x^{\mu=0,1,2}$ represents
imaginary time as well as spatial coordinates;
$  \bar{\psi} =
  \psi^{\dag}
  \gamma_{0}$;
and we have scaled out $v_{F}$.

Having identified
the low-energy fermionic degrees of freedom,
we now reinstate gauge field fluctuations.
The complete effective action is that of
SU(8) QED$_3$ coupled in addition with the Chern-Simons field 
$A^{\ }_{\mu}$ and given by
\begin{eqnarray}
&&
S_{\mathrm{eff}}
  =
\int d^3 x\,
\Biggr[
\sum_{(\alpha)=1}^8
  \bar{\psi}^{\ }_{(\alpha)}
\gamma_{\mu}
\left(
    \partial_\mu -{i} a_\mu-{i} A_\mu
  \right)\psi^{\ }_{(\alpha) }
\nonumber \\
&&
\quad
  +
  \mathcal{L}_{2f}
  +
  \mathcal{L}_{4f}
  +
  \frac{1}{2e^2}\left(
    \epsilon^{\ }_{\mu\nu\lambda}
    \partial^{\ }_{\nu}a^{\ }_{\lambda}
  \right)^2
  +
  \frac{{i}}{4\pi}
  \epsilon^{\ }_{\mu\nu\lambda}
  A^{\ }_{\mu}\partial^{\ }_{\nu}A^{\ }_{\lambda}
\Biggl] .
\nonumber \\
\label{eq: effective action}
\end{eqnarray}
Here we also included possible fermion bilinears
$\mathcal{L}_{2f}$ and four fermion interactions $\mathcal{L}_{4f}$.
$e^2 \sim U^{-1} $ stands for the coupling constant of the gauge field
$a_\mu$.

As discussed in
Refs.\ \onlinecite{Alicea05a, Alicea05b},
the Chern-Simons field $A_{\mu}$ is irrelevant.
The microscopic symmetries of the spin Hamiltonian
prohibit fermion mass terms $\mathcal{L}_{2f}$ from appearing
in the effective action (see Appendix \ref{sec : Fermion bilinears}).
Furthermore, four fermion interactions $\mathcal{L}_{4f}$ in
Eq.\ (\ref{eq: effective action})
are irrelevant when the number of fermion flavors
is large enough, $N>N_c$.
Thus, provided $N_c < 8$,
we have obtained a stable Algebraic Vortex Liquid phase that
is described by QED$_3$ with an emergent $\mathrm{SU}(8)$ symmetry.
Using the available theoretical understanding of such theories,
we will now describe the main properties of the AVL phase,
which in terms of the original spins is a gapless spin liquid phase.

\section{Properties of the AVL phase}
\label{sec : Properties of the AVL phase}

\subsection{Roton spectrum}
\label{sec:roton}

\begin{figure}
  \begin{center}
    \includegraphics[width=7cm,clip]{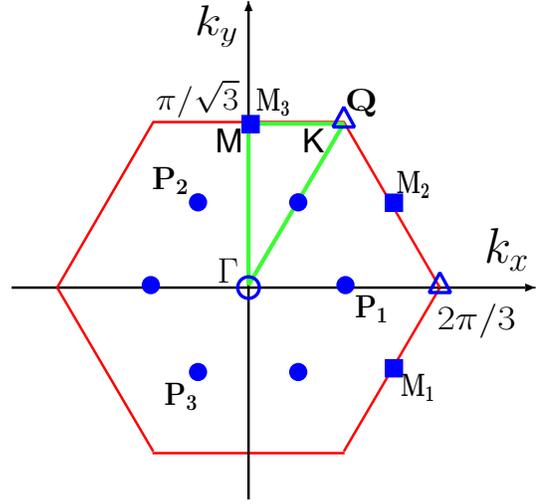}
\caption{
\label{fig: enhbilin.eps}
Summary of the main characterizations of the kagome AVL phase.
The low-energy excitations are located at wavevectors
${\bf 0}$, $\pm{\bf Q}$, ${\bf M}_{1,2,3}$, and $\pm{\bf P}_{1,2,3}$
in the physical Brillouin zone of the kagome lattice, where
${\bf Q} = (\pi/3, \pi/\sqrt{3})$;
${\bf M}_1 = (\pi/2, -\pi/(2\sqrt{3}))$,
${\bf M}_2 = (\pi/2, \pi/(2\sqrt{3}))$,
${\bf M}_3 = (0, \pi/\sqrt{3})$;
${\bf P}_1 = (\pi/3, 0)$,
${\bf P}_2 = (-\pi/6, \pi/(2\sqrt{3}) )$, and
${\bf P}_3 = (-\pi/6, -\pi/(2\sqrt{3}) )$.
Both $S^z$ and $S^+$ exhibit power law correlations at all these 
momenta, 
Eqs.~(\ref{eq: Sz structure factor})~and~(\ref{eq : S^+- structure factor}).  
The $S^z$ correlations are ``enhanced'' 
($\eta_z \approx 2.46$) for the subset
$\pm{\bf Q}$, $\pm{\bf P}_{1,2,3}$ and not enhanced 
($\eta_z = 3$)
for the remaining wavevectors.
On the other hand, all $S^+$ correlations are characterized
by the same exponent
$\eta_{\pm} \approx 3.24$.
}
\end{center}
\end{figure}

First note that the specific momenta of the Dirac points
Eq.~(\ref{eq: Dirac points}) and Fig.~\ref{fig: bz}
are gauge-dependent.
We are interested in the physical properties of the system, 
which are gauge invariant.
One such property is the vortex-antivortex excitation
spectrum:  By moving a vortex from an occupied state ${\bf k}$ 
with energy $E^- (\mathbf{k})$ to an unoccupied state 
$\mathbf{k'}$ with energy $E^+(\mathbf{k'})$,
we are creating an excitation that carries energy 
$E^+(\mathbf{k'}) - E^- (\mathbf{k})$.
Some care is needed if we want to construct such a state with
a definite momentum in the physical BZ, 
since we need to consider both $\mathbf{k'}-\mathbf{k}$ 
and $\mathbf{k'}-\mathbf{k} + \mathbf{G}_1$.
By examining Fig.~\ref{fig: bz},
we find that the vortex-antivortex continuum goes
to zero at twelve points
${\bf 0}$, $\pm {\bf Q}$, ${\bf M}_{1,2,3}$, $\pm {\bf P}_{1,2,3}$
in the BZ of the kagome lattice shown in
Fig.~\ref{fig: enhbilin.eps}.

\begin{figure}
  \includegraphics[width=\columnwidth]{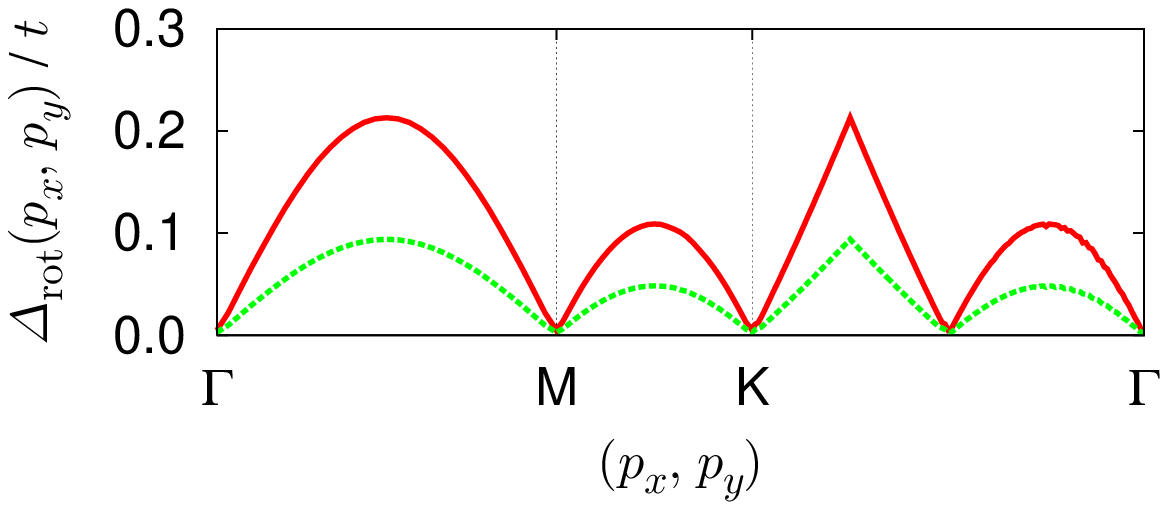} \\
  \includegraphics[width=\columnwidth]{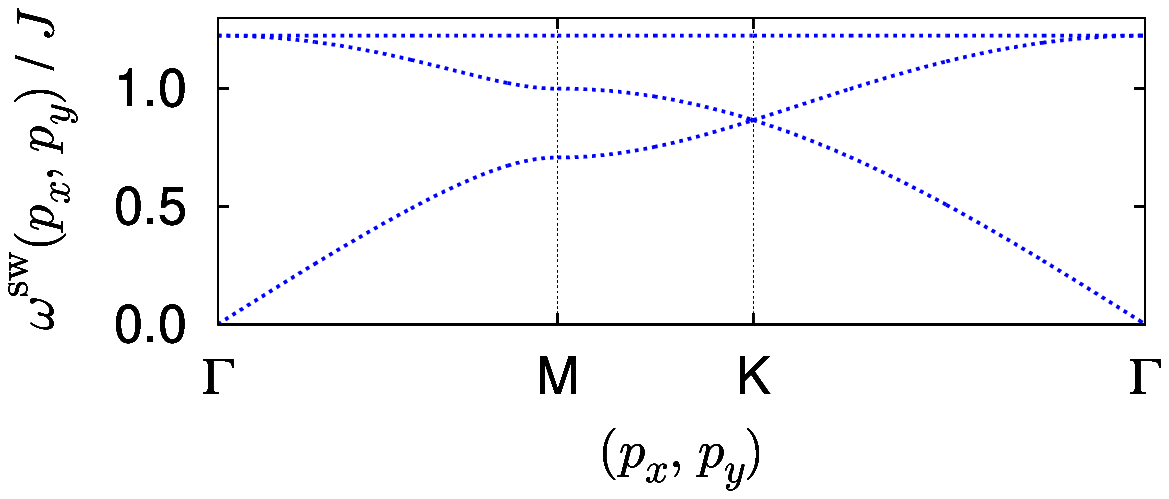}
  \caption{
    \label{fig:spinwaves}
    (Top) Lower edge of the vortex-antivortex continuum along several cuts
    in the Brillouin zone as shown in Fig.~\ref{fig: enhbilin.eps}
    for parameter values $t'/t=0.5$ and $t'/t=0.2$ from the higher to lower curve.
    (Bottom) Spin wave excitations, Eq.~(\ref{SW}), of the 
    kagome XY antiferromagnet with the nearest-neighbour exchange 
    $J$ only along the same cuts in the Brillouin zone.
  }
\end{figure}

An accompanying plot of the lower edge of the vortex-antivortex
continuum,
\begin{eqnarray}
\Delta_{\rm rot}(\mathbf{p}) &=&
\mathrm{min}^{\ }_{\mathbf{k}}\, 
\{ E^+(\mathbf{k} + \mathbf{p}) - E^- (\mathbf{k}) \},
\end{eqnarray}
which can be interpreted as a roton excitation energy,
is shown in Fig.\ \ref{fig:spinwaves} (top panel) along several BZ cuts.
Various proximate phases, e.g., magnetically ordered states,
will instead have gapped vortices.  But if such a gap is small, we expect
that the lower edge of the full excitation spectrum will be
dominated by deep roton minima near the same wavevectors (except where the rotons are masked by spin waves).

As a contrasting example, in Fig.~\ref{fig:spinwaves} (bottom panel)
we also show the spin-wave spectrum in the $q=0$ magnetically ordered 
phase\cite{Harris92,Chubukov92},
assuming this state is stabilized by some means.
(Actually, starting from {\em any} of the classically
degenerate ground states of the nearest-neighbor
kagome XY model leads to the same linear spin-wave
theory if one rotates the spin quantization axis suitably
for each site.)
  There are three branches.
For the XY antiferromagnet, these are
\begin{eqnarray}
\label{SW}
\omega^{\mathrm{sw}}_{1,2,3} (\mathbf{p}) 
&=& S J \sqrt{4 - \lambda_{1,2,3}(\mathbf{p})},
\nonumber \\
\lambda_1(\mathbf{p}) &=& -2, 
\nonumber \\
\lambda_{2,3}(\mathbf{p}) &=& 
1 \mp \sqrt{1 + 8 \cos p_1 \cos p_2 \cos p_3},
\end{eqnarray}
where $p_1 = p_x$, $p_2 =  p_x/2 + \sqrt{3}p_y/2$,
and $p_3 = p_x/2 - \sqrt{3}p_y/2$.
Clearly, the spin-wave frequency is large and of order $J$ throughout
the BZ except near zero momentum.  It is possible that higher order in $1/S$ spin wave corrections will lead to roton minima.  It would be interesting to
explore the possibility of such roton minima in the magnetically ordered
phase to see if they are coincident with the gapless momenta of the AVL phase as shown in Fig.~\ref{fig: enhbilin.eps}.  Such a coincidence between the roton minima 
obtained with spin wave perturbation theory\cite{Zheng06,Zheng06b,Starykh} 
and with the AVL approach\cite{Alicea05b} was indeed found within the magnetically ordered phase 
of the square lattice spin-1/2 quantum antiferromagnet with a weak second neighbor
exchange across one diagonal of each elementary square 
(a spatially anisotropic triangular lattice).
It would 
be useful to look for such signatures in the 
kagome experiments.

The momenta of the low-energy or gapless excitations capture important lattice-
to intermediate-scale physics and provide one of the simplest
characteristics that can be used to distinguish
between different spin liquid phases.
For example, in the algebraic spin liquid state proposed 
by Hastings and Ran\etal \cite{Hastings01, Ran06} 
for the $S=1/2$ nearest-neighbor Heisenberg model,
the gapless spinon-antispinon excitations occur
at four momenta
\{$\mathbf{0}$, $\mathbf{M}_{1,2,3}$\}
in the BZ.
On the other hand, Sachdev\cite{Sachdev91} and
Wang and Vishwanath \cite{Wang06}
discuss four different gapped $Z_2$ spin liquid
phases on the kagome lattice using Schwinger bosons.
One of the states has minimum in the gapped spin excitations at $\mathbf{0}$; 
one has minima at $\pm \mathbf{Q}$; and the two remaining
spin liquids have minima at
the same 12 momenta as the AVL phase, Fig.~\ref{fig: enhbilin.eps}.

\subsection{
$S^{z} S^{z}$ correlator
}
\label{sec: vortex-antivortex excitation and Sz correlator }

A more formal approach towards characterizing the system with
gauge interactions is to deduce
the PSG transformations
for a particular mean-field state.\cite{Wen02}  
Specifically, the notion of symmetry is enlarged to also include gauge transformations
in order to maintain invariance of the mean-field
Hamiltonian under the usual symmetry operations.

Some details are presented in Appendix~\ref{sec : Fermion bilinears},
where we show the calculated transformation properties of
the continuum fermion fields $\bar{\psi},\psi$.
In particular, with the PSG analysis, the low-energy roton excitations
can be studied by analyzing fermion bilinears.

The $S^z$-component of the spin is expressed as
a gauge flux in our QED$_3$ theory of vortices.
Both the flux induced by
the dynamical U(1) gauge field itself
and the flux induced by vortex currents
contribute to the $S^z S^z$ correlation.
Formally,
the $S^{z}$ operator can be expressed
in terms of the gauge flux and
appropriate fermion bilinears
$\bar{\psi} {\cal G} \psi $
representing the contributions by vortex currents,
\begin{eqnarray}
S^{z}_{\mathbf{r}}
&\sim  &
\frac{\nabla\times \boldsymbol{a}}{2\pi}
+
\sum_{i}
A_i
e^{-i \mathbf{K}_i\cdot \mathbf{r}}
\bar{\psi} {\cal G}_i \psi
+
\cdots
\label{eq: Sz and bilinears}
\end{eqnarray}
where the summation over $i$ runs over
bilinears $\bar{\psi} {\cal G}_i \psi$
with appropriate symmetry properties so that the 
right hand side transforms identically to $S^{z}$.
The factor
$e^{- i \mathbf{K}_i\cdot \mathbf{r}}$
represents the momenta carried by
these bilinears,
which are the same as the roton minima shown in
Fig.\ \ref{fig: enhbilin.eps}, 
and $A_i$ is some amplitude for the bilinear ${\cal G}_i$
[in fact, one needs to specify four such amplitudes $A_i(j)$ since there
are four sites $j=1,\ldots,4$ in the unit cell, 
Fig.\ \ref{fig: kagome_lattice}].

Assuming the $N=8$ QED$_3$ is critical, all correlation functions
decay algebraically.
We then expect to find power law correlation at all momenta
shown in Fig.~\ref{fig: enhbilin.eps}.
As a consequence,
the singular part of the structure factor
at zero temperature near such wavevector ${\bf K}$ is given by
\begin{eqnarray}
S^{zz}(\mathbf{k}=\mathbf{K}+\mathbf{q},\omega)
\propto
\frac{\Theta(\omega^2-\mathbf{q}^2)}{(\omega^2-\mathbf{q}^2)^{(2-\eta_z)/2}}.
\label{eq: Sz structure factor}
\end{eqnarray}
The exponent $\eta_z$ characterizes the strength of the $S^z$ correlations,
and smaller values correspond to more singular and therefore more
pronounced correlations.
In QED$_3$, the corresponding exponent value for the flux 
$\nabla \times a$ that contributes to the $S^z$ correlation near
zero momentum is $\eta_z = 3$, and the same exponent also characterizes
fermionic bilinears that are conserved currents, 
see Appendix~\ref{sec : Fermion bilinears}.
On the other hand, bilinears that are not conserved currents
have their scaling dimensions enhanced by the gauge field fluctuations
as compared with the mean field.
Upon analyzing the transformation properties of the fermionic bilinears,
we conclude that there are indeed such enhanced contributions to
$S^z$ near the momenta $\pm {\bf Q}$ and $\pm {\bf P}_{1,2,3}$
(explicit expressions are given in Appendix~\ref{sec : Fermion bilinears}).
The new exponent can be estimated from the large-$N$ treatments
of the QED$_3$\cite{Rantner02, Franz03}
to be $\eta_z \approx 3 - 128/(3\pi^2 N)  = 2.46$
for $N=8$.

\subsection{$S^+ S^-$ correlator}
\label{sec: S+S- correlator}

Since $S^z \sim (\nabla \times a)/(2\pi)$, the spin raising 
operator $S^+$ in the dual theory is realized as an operator that 
creates $2\pi$ gauge flux.  Following Ref.~\onlinecite{Alicea05b},
we treat such flux insertion (``monopole insertion'')
classically as a change of the background gauge field configuration
felt by fermions. 
By determining quantum numbers carried by the monopole insertion 
operators, we can find the wavevectors of the dominant 
$S^+$ correlations.

Here we only outline the procedure 
(for details, see Refs.~\onlinecite{Alicea05, Alicea05b}).
In the presence of $2\pi$ background flux inserted at $\mathbf{x}=0$,
each Dirac fermion species $(\alpha) = 1\dots 8$ has a quasi-localized 
zero-energy state near the inserted flux.
We will denote the creation operators for such fermionic zero
modes by $f^{\dag}_{(\alpha)}$.
The corresponding two-component wavefunctions have the form
$\varphi(\mathbf{x})
\sim \frac{1}{|\mathbf{x}|}
  \left(
    \begin{array}{c}
      1 \\
      0
    \end{array}
  \right)
$,
which allows us to relate the transformation properties of
$f^{\dag}_{(\alpha)}$ to those of the first component of 
$\psi^{\dagger}_{(\alpha)}$ (the latter are summarized in 
Appendix~\ref{app : PSG}).

The monopole creation operators ${\cal M}^\dag$ 
are defined by the 
combination of the flux insertion and the subsequent filling of 
four out of eight fermionic zero modes in order to ensure 
gauge neutrality:
\begin{eqnarray}
{\cal M}^{\dag}|DS_0 \rangle
=
f^{\dag}_{({\alpha})}
f^{\dag}_{({\beta})} 
f^{\dag}_{({\gamma})} 
f^{\dag}_{({\delta})} 
|DS_+\rangle,
\end{eqnarray}
where $|DS_0 \rangle$ and $|DS_+\rangle$ are the Fermi-Dirac sea in the 
absence and presence of $+2\pi$ flux, respectively.
There are total of $70$ distinct such monopoles.
The quantum numbers of the monopole operators
are then determined by the transformation properties
of the operators $
f^{\dag}_{(\alpha) }
f^{\dag}_{(\beta) } 
f^{\dag}_{(\gamma) } 
f^{\dag}_{(\delta) }$
and those of 
$|DS_+ \rangle$
relative to $|DS_0 \rangle$.

Both $|DS_+ \rangle$ and 
$|DS_0 \rangle$ 
are expected to be eigenstates of the translation
operator $T_{\delta\mathbf{r}}$, and we denote the ratio of the 
eigenvalues as 
$e^{{i}\mathbf{\Pi}_{\rm offset} \cdot \delta \mathbf{r}}$.
On the other hand, we can diagonalize the action of 
$T_{\delta\mathbf{r}}$ on
$
f^{\dag}_{(\alpha) } 
f^{\dag}_{(\beta) } 
f^{\dag}_{(\gamma) } 
f^{\dag}_{(\delta) }
$
to get a set of eigenvalues
$e^{{i}\mathbf{\Pi}_i \cdot \delta \mathbf{r}}$, $i=1,\dots,70$.
Thus,
the momenta of the 
monopole operators (at zero energy) 
are determined by $\{ \mathbf{\Pi}_i \}$ which are offset by the
hitherto unknown wavevector $\mathbf{\Pi}_{\rm offset}$.
However, once we examine $\{ \mathbf{\Pi}_i \}$, 
we find that the offset 
is in fact uniquely fixed by the requirement that the 
monopole momenta are distributed in the physical BZ
in a way that respects all lattice symmetries.

Monopole momenta determined in this manner are found to be 
given by the very same wavevectors
${\bf 0}$, 
$\pm{\bf Q}$,
${\bf M}_{1,2,3}$, and
$\pm{\bf P}_{1,2,3}$, as the leading $S^z$ correlation
(Fig.\ \ref{fig: enhbilin.eps}).
For the record, we find the following monopole multiplicities given in 
the parentheses: ${\bf 0}$~(12 monopoles), $\pm{\bf Q}$~(5 each),
${\bf M}_{1,2,3}$~(8 each), and $\pm{\bf P}_{1,2,3}$~(4 each).

The monopole insertions which add $S^z=1$
have overlap with the exact lowest energy spin carrying eigenstates in the AVL. 
At the lowest energies such eigenstates will be centered around the monopole  wavevectors in the Brillouin zone, and we expect
that they will disperse with an energy growing linearly with small
deviations in the momenta.   But presumably any local operator that adds spin one
will create a linearly dispersing but overdamped ``particle'' excitation.
More formally, 
we can expand the $S^+$ operator in terms of the 
continuum fields as follows,
\begin{eqnarray}
S^+_{\mathbf{r}}
&\sim&
\sum_i
A_i 
e^{{i}\mathbf{\Pi}_i\cdot \mathbf{r}}
{\cal M}^{\dag}_i,
\label{eq: S+ and monopoles}
\end{eqnarray}
where $\mathbf{\Pi}_i$ is the momentum of the monopole ${\cal M}_i$.
The right hand side has correct transformation properties under
the translations, but not all of the monopole operators
contribute to $S^+$ when other symmetries are taken
into account.  For example, some of the zero momentum monopoles have
opposite eigenvalues under the lattice inversion, and the same
happens for the momenta ${\bf M}_{1,2,3}$.
However, given the large monopole multiplicities, 
it appears very likely that all momenta shown in 
Fig.~\ref{fig: enhbilin.eps} will be present in $S^+$ 
(but we have not performed an analysis of all quantum numbers so far).

The in-plane dynamical structure factor $S^{+-}$ around each 
${\bf \Pi}_i$ is thus expected to have the same critical form as $S^{zz}$, 
\begin{eqnarray}
S^{+-}(\mathbf{k}=\mathbf{\Pi}+\mathbf{q},\omega)
\propto
\frac{\Theta(\omega^2-\mathbf{q}^2)}{(\omega^2-\mathbf{q}^2)^{(2-\eta_{\mathrm{\pm}})/2}} ~,
\label{eq : S^+- structure factor}
\end{eqnarray}
but the characteristic exponent $\eta_{\pm}$ is now that appropriate 
for the monopole operators and is expected to be the same for all 
momenta.
Ref.\ \onlinecite{Borokhov02} calculated this in the large-$N$ 
QED$_{3}$: $\eta_{\mathrm{mon}} \approx 0.53 N - 1$.
Setting $N=8$, we estimate 
$\eta_{\mathrm{\pm}} \approx 3.24$, which is larger than even the
non-enhanced exponent $\eta_z = 3$ entering the $S^z$ correlations.

The above results with  $\eta_{\pm} > \eta_z$ suggest that within the AVL phase
the system is ``closer'' to an Ising ordering of $S^z$ than of an in-plane XY ordering of $S^+$ .   This is rather puzzling for an easy-plane spin model,
where one would expect  XY order to develop more readily than Ising order.
This conclusion is reminiscent of the
classical kagome spin model, where the easy-plane anisotropy present in the XY model destroys the zero temperature coplanar order of the Heisenberg model.
However, 
this interpretation of the inequality $\eta_{\pm} > \eta_z$ is perhaps somewhat misleading.  Indeed, as we discuss in Appendix~\ref{sec : Fermion bilinears},
condensation of any of the enhanced fermionic bilinears that 
contribute to $S^z$ would drive XY order in addition to  $S^z$ order - forming a supersolid phase of the bosonic spins.
Thus, care must be taken when translating the long-distance behavior of the AVL correlators into ordering tendencies.

The exact diagonalization study by Sindzingre\cite{Sindzingre} finds 
that in the nearest-neighbor kagome XY antiferromagnet the $S^+$ 
correlations are larger than the $S^z$ correlations, 
which is different from the AVL prediction.  
It would be interesting to examine this further, perhaps also in a 
model with further-neighbor exchanges.
With regards to such numerical studies, we also want to point out that 
the discussed dynamical spin structure factor, 
Eqs.~(\ref{eq: Sz structure factor})~and~(\ref{eq : S^+- structure factor}),  
translate into equal-time spin correlations that decay as 
$r^{-1-\eta_{z,\pm}}$.
For the estimated values of $\eta_{z,\pm}$, the decay is rather quick, 
and could be hard to distinguish from short-range correlations.

\section{Conclusion}
\label{sec: conclusion}

Using fermionized vortices,
we have studied the 
easy-plane spin-1/2 Heisenberg antiferromagnet on the 
kagome lattice
and accessed a novel gapless spin liquid phase.
The effective field theory for the resulting Algebraic
Vortex Liquid phase, is (2+1)D QED$_3$ with an emergent $\mathrm{SU}(8)$ 
flavor symmetry.
This large number of flavors 
can be thought to have its origin in
many competing ordered states that are intrinsic to the ``loose'' network
of the corner-sharing triangles in the kagome lattice.
It is also likely that for this number of Dirac fermions the
critical QED$_3$ theory is stable against the dynamical gap formation,
which allows us to expect that the AVL phase is stable.

The gapless nature of the AVL phase has important thermodynamic consequences.
For example, we predict the specific heat to behave as $C \sim T^2$
at low temperatures.  Since vortices carry no spin, we expect this 
contribution to be unaffected by the application of a magnetic field.  
Interestingly, such behavior was observed in 
SrCr${ }_{8-x}$Ga${}_{4+x}$O$_{19}$ and was interpreted in terms of 
singlets dominating the many-body spectrum.\cite{Ramirez00, Sindzingre00}
Since the vortices are mobile and carry energy, we furthermore predict 
significant thermal conductivity, 
which would be interesting to
measure in such candidate gapless spin liquids.

The more detailed properties of the kagome AVL phase were studied 
by the PSG analysis, leading to detailed predictions for the spin structure
factor.
We found that the dominant low energy spin correlations occur at twelve
specific locations in the BZ shown in Fig.\ \ref{fig: enhbilin.eps}.
These wavevectors encode important intermediate-scale physics and can be 
looked for in experiments and numerical studies.
They may be also used to 
compare and contrast with other 
theoretical proposals of spin liquid states.

The recent experiments\cite{Helton06,Ofer06,Mendels06} 
for the spin-1/2 kagome material ZnCu$_3$(OH)$_6$Cl$_2$
observed power-law behavior in $\omega$ of the structure factor 
$S(\mathbf{k},\omega)$ in a powder sample.
The specific heat measurements and local magnetic probes also 
suggest gapless spin liquid physics in this compound.
Unfortunately, no detailed momentum-resolved information is 
available so far.
It should be also noted that our results were derived assuming
easy-plane character of the spin interactions, and do not apply
directly if the material has no significant such spin anisotropy.
\cite{remark 2}
Hopefully, more experiments in the near future will provide detailed 
microscopic characterization of this material,
such as the presence/absence of the
Dzyaloshinskii-Moriya interaction \cite{Rigol07},
and clarify the appropriate spin modeling.
In the case of the Heisenberg spin symmetry and nearest-neighbor
exchanges, Refs.~\onlinecite{Hastings01, Ran06} offer another
candidate critical spin liquid for the kagome lattice, obtained using a
slave fermion construction.
This algebraic spin liquid (ASL) state differs significantly from the presented AVL phase
as discussed in some detail at the end of the introduction and towards
the end of Sec.~\ref{sec:roton}.  
Briefly, 
within the ASL the momentum space locations of the gapless
spin carrying excitations are only a small subset of those predicted for the AVL. 
Moreover, 
while both phases support gapless Dirac fermions that contribute a $T^2$ specific heat, in the AVL phase the fermions (vortices) are spinless
in contrast to the fermionic spinons in the ASL.  Thus, one would expect
the specific heat to be more sensitive to an applied magnetic field in the ASL than in the AVL.

Contrary to SrCr$_{8-x}$Ga$_{4+x}$O$_{19}$, 
the experimental data on ZnCu$_3$(OH)$_6$Cl$_2$
shows that the specific heat is affected by the magnetic field
in the low temperature regime \cite{Helton06}.
However, as pointed out by Ran \etal,\cite{Ran06} 
$T \sim 10\mathrm{K}$ is likely to be
an appropriate temperature scale to test several
theoretical
approaches to this material, since spin liquid physics
might be masked by, for example, impurity effects,
Dzyaloshinskii-Moriya
coupling, or other complicating spin interactions,
at the lowest temperature scales. (Indeed, the
susceptibility data
are perhaps consistent with the presence of impurities,
although it is unclear at this stage if the peculiar
temperature dependence of the specific heat
also has its origins in impurities.)
The theoretical prediction of the $T^2$ specific heat is
consistent
with the experiments for $T > 10 \mathrm{K}$.

A general outstanding issue is the connection, if any,
between the critical spin liquids obtained with slave fermions and with
fermionic vortices.  Perhaps our AVL phase corresponds to some algebraic spin liquid ansatz, 
but at the moment this is unclear.

\section*{Acknowledgments}
We would like to thank P.~Sindzingre for sharing his exact
diagonalization results on the spin-1/2 XY kagome system.
This research was supported 
by the National Science Foundation through grants
PHY99-07949 and DMR-0529399 (M.\ P.\ A.\ F and J.\ A.\ ).

\appendix

\section{Zero mode wave functions at the Fermi points}
\label{sec : Zero mode wave functions at the Fermi points}

In this appendix, the 16-component wavefunctions
$\phi^{\ }_{i,\alpha,a}(n)$ ($n=1,\ldots,16$)
representing zero modes at the Fermi points
are presented explicitly.
These wave functions are necessary for
the PSG analysis.

At the node $\mathbf{Q}_1$,
\begin{eqnarray}
  \phi^{\ }_{\mathrm{I},1,1}&=&
  e^{{i} \pi/4}
  \left(
    \begin{array}{c}
      \chi^{\mathbf{Q}_1}_{A+} \\
      0
    \end{array}
  \right),\quad
  \phi^{\ }_{\mathrm{I},1,2}=
  e^{{i} \pi/12}
  \left(
    \begin{array}{c}
      \chi^{\mathbf{Q}_1}_{A-} \\
      0
    \end{array}
  \right),\quad
\nonumber \\
  \phi^{\ }_{\mathrm{II},1,1}&=&
  e^{-{i}\pi/4}
  \left(
    \begin{array}{c}
      0 \\
      \chi^{\mathbf{Q}_1}_{B+}
    \end{array}
  \right),\,
  \phi^{\ }_{\mathrm{II},1,2}=
  e^{-{i} \pi/12}
  \left(
    \begin{array}{c}
      0 \\
      \chi^{\mathbf{Q}_1}_{B-}
    \end{array}
  \right) ~;
\nonumber \\
\end{eqnarray}
at the node $\mathbf{Q}_2$,
\begin{eqnarray}
  \phi^{\ }_{\mathrm{I},2,1}&=&
  \left(
    \begin{array}{c}
      \chi_{A+}^{\mathbf{Q}_2} \\
      0
    \end{array}
  \right),\quad
\phi^{\ }_{\mathrm{I},2,2}=
  e^{{i} \pi/6}
  \left(
    \begin{array}{c}
      \chi_{A-}^{\mathbf{Q}_2} \\
      0
    \end{array}
  \right),
\nonumber \\
\phi^{\ }_{\mathrm{II},2,1}&=&
  \left(
    \begin{array}{c}
      0 \\
      \chi_{B-}^{\mathbf{Q}_2}
    \end{array}
  \right),\quad
 \phi^{\ }_{\mathrm{II},2,2}=
  e^{-{i} \pi/6 }
  \left(
    \begin{array}{c}
      0 \\
      \chi_{B+}^{\mathbf{Q}_2}
    \end{array}
  \right);
\nonumber \\
\end{eqnarray}
at the node $\mathbf{Q}_3$,
\begin{eqnarray}
\phi^{\ }_{\mathrm{I},3,1}&=&
  e^{{i} 5\pi/4}
  \left(
    \begin{array}{c}
      0 \\
      \chi_{B-}^{\mathbf{Q}_3}
    \end{array}
  \right),\quad
  \phi^{\ }_{\mathrm{I},3,2}=
  e^{{i} 13 \pi/12 }
  \left(
    \begin{array}{c}
      0 \\
      \chi_{B+}^{\mathbf{Q}_3}
    \end{array}
  \right),
\nonumber \\
  \phi^{\ }_{\mathrm{II},3,1}&=&
  e^{-{i} \pi/4}
  \left(
    \begin{array}{c}
      \chi_{A+}^{\mathbf{Q}_3} \\
      0
    \end{array}
  \right),\,
  \phi^{\ }_{\mathrm{II},3,2}=
  e^{-{i} \pi/12 }
  \left(
    \begin{array}{c}
      \chi_{A-}^{\mathbf{Q}_3} \\
      0
    \end{array}
  \right);
\nonumber \\
\end{eqnarray}
and at the node $\mathbf{Q}_4$,
\begin{eqnarray}
  \phi^{\ }_{\mathrm{I},4,1}&=&
  -
  \left(
    \begin{array}{c}
      0 \\
      \chi_{B+}^{\mathbf{Q}_4}
    \end{array}
  \right),\quad
  \phi^{\ }_{\mathrm{I},4,2}=
  e^{{i} 7 \pi/6}
  \left(
    \begin{array}{c}
      0 \\
      \chi_{B-}^{\mathbf{Q}_4}
    \end{array}
  \right),\quad
\nonumber \\
  \phi^{\ }_{\mathrm{II},4,1} &=&
  \left(
    \begin{array}{c}
      \chi_{A+}^{\mathbf{Q}_4} \\
      0
    \end{array}
  \right),\quad
  \phi^{\ }_{\mathrm{II},4,2}=
  e^{-{i} \pi/6 }
  \left(
    \begin{array}{c}
      \chi_{A-}^{\mathbf{Q}_4} \\
      0
    \end{array}
  \right);
\nonumber \\
\end{eqnarray}
where a convenient choice for
the eight-component zero energy wave functions $\chi_A$ and $\chi_B$ is
\begin{equation}
  \chi_{A,s}^{\mathbf{Q}_{\alpha}}
  =
  N_A
  \left(
    \begin{array}{c}
      1 \\
      \alpha \\
      (x y)^2 (y^4 + (x y^*)^{2}) \\
      -\alpha (x y)^2 (y^4 -(x y^*)^{2}) \\
      - x y^{*3} + \alpha x y^* (y^4 - (x y^*)^{2}) \\
      t'
      \left[
        x^* y^{*3} +\alpha x y^{*3}
      \right] \\
       - x^3 y^* \left(y^4 + (x y^*)^{2}\right)
      -\alpha  x^* y^{*3}  \\
      t' x^{3} y^*
      \left[
y^4 x^{* 2}
         + y^{*2}
-        \alpha
        \left(
          y^4-(x y^*)^{2}
        \right)
      \right]
    \end{array}
  \right),
\end{equation}
\begin{equation}
  \chi_{B,s}^{\mathbf{Q}_{\alpha}}
  =
  N_B
  \left(
    \begin{array}{c}
      1 \\
      \beta        \\
      (x^{*} y^{*})^2 (y^{*4} + (x^{*} y)^{2}) \\
      -\beta (x^{*} y^{*})^2 (y^{*4} - (x^{*} y)^{2}) \\
            - t' \left[
         x^{*} y^{3} + \beta  x y^3
      \right]\\
            x^* y
      \left(
        y^{*4}+ (x^{*}y)^2
      \right)
      +\beta x^{*} y^{3}
      \\
        t' x^{*3} y\left[
        \beta   \left(  
        y^{*4} x^{2}
           -y^2        \right)
        -  y^{*4} - (x^{*} y)^{2}
      \right]
      \\
          x y^3
      -
      \beta
      x^{*3} y \left(
        y^{*4}- (x^{*} y)^{2}
      \right)
    \end{array}
  \right).
\end{equation}
Here
$x := e^{{i} k_x/2}$, $y :=
e^{{i} k_y/(2\sqrt{3})}$ for ${\bf Q} = (k_x, k_y)$,
\begin{eqnarray}
      \alpha &=&
      \frac{
              s 3 \sqrt{2} x^{*3}y^{*3}
        -(1+x^6 y^6)(x^4 + x^{*4})
}{2 x^{-2}(3- x^2 y^{* 6} -x^{*2}y^6)},
\nonumber \\
      \beta &=&
      \frac{
         s {i} 3 \sqrt{2} x^{*3}y^{*3}
        -(1-x^6 y^6)(x^4 + x^{*4})
}{2 x^{-2}(3- x^2 y^{* 6} -x^{*2}y^6)},
\end{eqnarray}
and $N_A$ and $N_B$ are $t'$-dependent normalization factors. 
These zero modes are chosen in such a way that the $s =\pm 1$
wavefunctions are orthogonal to each other for any $t'$.

In order to work with the continuum Dirac fields 
Eq.~(\ref{eq: def continuum field}),
it is convenient to introduce four sets of
the Pauli matrices which act on different gradings.
In the following,
Pauli matrices denoted by $\rho^{\kappa}$ ($\kappa=0,1,2,3$)
act on Latin indices $a=1,2$
($\rho^{0}$ is an identity matrix).
Pauli matrices denoted by
$\sigma^{\kappa}$ (and hence $\gamma^{\kappa}$
introduced in Eq.~\ref{eq: gamma matrices})
 act on Latin indices  $i=\mathrm{I},\mathrm{II}$.
We separate the node momenta into
two groups $(\mathbf{Q}_1,  \mathbf{Q}_3)$
and $(\mathbf{Q}_2,\mathbf{Q}_4)$. 
Then, Pauli  matrices denoted by $\tau^{\kappa}$
act within a given group, either
  $(\mathbf{Q}_1, \mathbf{Q}_3)$ or $(\mathbf{Q}_2,\mathbf{Q}_4)$.
Finally, Pauli matrices denoted by $\mu^{\kappa}$ act on
the space spanned by
  $(\mathbf{Q}_{1,3}, \mathbf{Q}_{2,4})$.
We will also use the notation
$\tau^{\pm}=\tau^1 \pm i \tau^2$.

\section{Projective symmetry group (PSG) analysis}
\label{app : PSG}

Any symmetry of the original lattice spin model has
a representation in terms of vortices.
Unfortunately, upon fermionization,
time-reversal symmetry transformation
becomes highly non-local
and we do not know how to implement
it in the low-energy effective field theory.

Since a specific configuration
of the Chern-Simons gauge field was picked
in the flux smearing mean field,
we had to fix the gauge and work with the enlarged unit cell.
The spatial symmetries of the original lattice spin model are,
however, still maintained if the symmetry transformations are
followed by subsequent gauge transformations --
the original symmetries
become ``projective symmetries'' in the effective field theory.
The PSG analysis is necessary
when we try to make a connection
between the original spin model and the effective
field theory.

The original spin system has the following symmetries besides
the global $\mathrm{U}(1)$: 
\begin{eqnarray}
  &&
  T_{\mathbf{E}_{1}/2}:\quad \mbox{translation by $\mathbf{E}_{1}/2$},
  \nonumber \\
  &&
  T_{\mathbf{E}_{2}}:\quad  \mbox{translation by $\mathbf{E}_{2}$},
  \nonumber \\
  &&
  R_{\pi}:\quad  \mbox{rotation by $\pi$ around a kagome site},
  \nonumber \\
  &&
  \mathcal{R}_{x}: \quad  \mbox{reflection with respect to $y$ axis},
  \nonumber \\
  &&
  \mathcal{C}: \quad  \mbox{particle-hole},
  \nonumber \\
  &&
  \mathcal{T}: \quad  \mbox{time-reversal}.
\end{eqnarray}
We can also include a rotation by $\pi/3$ about honeycomb center but
will not consider it here.

The first four transformations
$T_{\delta  \mathbf{r}}$, $R_{\theta}$, $\mathcal{R}_{x}$ act on spatial
coordinates $\mathbf{r}=(x,y)$,
$\mathbf{r}\to \mathbf{r}+\delta \mathbf{r}$,
$\mathbf{r}\to (x \cos\theta  -y \sin\theta , x \sin\theta  + y \cos\theta )$,
$\mathbf{r}\to (-x,y)$,
respectively.
On the other hand,
the particle-hole symmetry $\mathcal{C}$
and the time reversal $\mathcal{T}$
act on spin operators as
\begin{eqnarray}
  \mathcal{C}: &&
  S^{z} \to -S^{z},
  \quad
  S^{x}\pm {i}S^{y} \to
  S^{x}\mp {i}S^{y},
\nonumber \\
  \mathcal{T}: &&
  \mathbf{S} \to - \mathbf{S},
  \quad
  {i}\to -{i}.
\end{eqnarray}
Here, the anti-unitary nature of the time reversal is
reflected in its action on the complex number ${i}\to -{i}$.

Symmetries in terms of the rotor representation can then be deduced as
\begin{eqnarray}
  \mathcal{C}:
  &\quad&
  n\to -n,\quad
  \varphi \to -\varphi,
  \nonumber \\
  \mathcal{T}:
  &\quad&
  n\to -n,\quad
  \varphi \to \varphi+\pi, \quad
  {i}\to -{i}.
\end{eqnarray}

Symmetry properties of
bosonic and fermionic vortices are deduced from
their defining relations.
Due to the Chern-Simons flux attachment,
the 
mirror and time-reversal symmetries are
 implemented in a non-local
fashion in the fermionized theory.
However,
if we combine $\mathcal{T}$ with
$\mathcal{R}_{x}$ and $\mathcal{C}$, the resulting modified reflection
$\tilde{\mathcal{R}}_{x} := \mathcal{R}_{x}\mathcal{C}\mathcal{T}$
can still be realized locally.
We summarize the symmetry properties of the fermions $d,d^{\dag}$
in Table \ref{tab: Summary of symmetry transformations}.
We also introduce a formal fermion time reversal by
\begin{eqnarray}
  \mathcal{T}_{\mathrm{ferm}}
  :
  \quad
  d\to d,
  \quad
  {i}\to -{i}.
\end{eqnarray}
The necessary explanations are the same as in 
Refs.~\onlinecite{Alicea05, Alicea05b} and are not repeated here.

\begin{table*}
  \begin{center}
    \begin{tabular}{c|c|c}
      & spin $\mathbf{S}$ & rotor $\varphi$, $n$   \\ \hline\hline
      $\mathcal{C}$ &
      $S^{y}\to -S^{y}$, $S^{z}\to -S^{z}$ &
      $n\to -n$, $\varphi \to -\varphi$                                
      \\ \hline
      $\mathcal{T}$ &
      $\mathbf{S}\to -\mathbf{S}$, ${i}\to -{i}$ &
      $n\to -n$, $\varphi \to \varphi +\pi$, ${i}\to -{i}$  
      \\ \hline
      $\tilde{\mathcal{R}}_{x}$ &
      $S^{x}\to -S^{x}$, ${i}\to -{i}$, $x\to -x$ &
      $n\to +n$, $\varphi \to -\varphi +\pi$, ${i}\to -{i}$, $x\to -x$
      \\
    \end{tabular}
  \end{center}
  \begin{center}
    \begin{tabular}{c|c|c}
      & vortex $\theta$,$N$ and gauge field $a$, $e$ & fermion $d,d^{\dag}$\\ \hline\hline
      $\mathcal{C}$ &
      $a \to -a$, $\theta\to -\theta$, $e\to -e$, $N\to 1-N$ &
      $d\to (-1)^i d^\dag$ \\ \hline
      $\mathcal{T}$ &
      $a \to -a$, $\theta\to -\theta$, $e\to e$, $N\to N$, ${i}\to -{i}$
      \\ \hline
      $\tilde{\mathcal{R}}_{x}$ &
      $a \to -a$, $\theta\to -\theta$, $e\to e$, $N\to N$, ${i}\to -{i}$,
      $x\to -x$ &
      $d\to d$, $x\to -x$, ${i}\to -{i}$
      \\
    \end{tabular}
  \end{center}
  \caption{
\label{tab: Summary of symmetry transformations}
Summary of symmetry transformations
for spin, rotor, vortex, and fermionized vortex operators on the
 lattice.
The sign factor $(-1)^{i}$ in
the action of the particle-hole transformation
on the fermionized vortices,
$\mathcal{C} : d\to (-1)^i d^\dag$,
is $1$ on one of the sublattice of the dual lattice
whereas it is $-1$ on the other.
}
\end{table*}

Finally, the symmetry properties of the slowly varying continuum fields $\bar{\psi}, \psi$
can be deduced from Eq.\ (\ref{eq: def continuum field}).
As explained, these are realized projectively,\cite{Wen02}
and the symmetry transformations for the continuum fermion fields
are summarized in 
Table \ref{tab: Summary of symmetry transformations for cont fermions}.

\begin{table*}
\begin{center}
\begin{tabular}{c|c|c|c|c|c|c}
 & $T_{\mathbf{E}_1/2}$ &$T_{\mathbf{E}_2}$  & $\tilde{\mathcal{R}}_{x}$
 & $R_{\pi}$ & $\mathcal{C}$ & $\mathcal{T}_{\mathrm{fermi}}$  \\ \hline \hline
$\psi \quad \to \hphantom{AA}$ &
  $\mu^{2}
  e^{-{i} \pi \tau^3/3}
   \rho^{1}
  \psi$
&
  $\mu^{3} e^{{i} \pi\tau^{3}/6}
  \psi $
&
 $
e^{+ {i} \pi\tau^3/12}
\tau^3
e^{+ {i} \pi\mu^2\rho^1/4}
\rho^3
\psi
 $
&
$\sigma^3 \rho^2
\tau^1
e^{{i} 5\pi\tau^3/12}
\mu^3
\psi$
&
$\tau^1\sigma^1 \left[\psi^{\dag}\right]^{T}$
&
$\tau^2 \sigma^2 \psi$
\end{tabular}
  \caption{
\label{tab: Summary of symmetry transformations for cont fermions}
Summary of symmetry transformations for continuum fermion fields
(upto unimportant $\mathrm{U}(1)$ phase factors).}
\end{center}
\end{table*}

\section{Fermion bilinears}
\label{sec : Fermion bilinears}

Once we determine the symmetry properties
of the continuum fermion fields $\bar{\psi},\psi$,
we can discuss symmetry properties of
the gauge invariant  bilinears
$\bar{\psi}\mathcal{G}\psi$
where $\mathcal{G}$ is a $16 \times 16$ matrix.
Using the gradings specified at the end of 
Appendix~\ref{sec : Zero mode wave functions at the Fermi points},
the bilinears can be conveniently written as
\begin{eqnarray}
  \bar{\psi}
  \sigma^{\kappa} \mu^{\lambda} \tau^{\alpha} \rho^{\beta}
  \psi
\end{eqnarray}
where $\kappa,\lambda,\alpha,\beta$ run from $0$ to $3$
and hence there are $4^{4}=64\times 4$ such bilinears. 
Among these bilinears,
$64\times 3$ bilinears with the Lorentz index $\kappa=1,2,3$
represent $\mathrm{U}(8)$ currents,
\begin{eqnarray}
  \bar{\psi}
  \sigma^{1,2,3} \mu^{\lambda} \tau^{\alpha}
  \rho^{\beta}
  \psi.
\label{eq: current bilinears}
\end{eqnarray}
On the other hand,
there are $64$ remaining bilinears with $\kappa=0$,
\begin{eqnarray}
  \bar{\psi}
  \sigma^{0} \mu^{\lambda} \tau^{\alpha} \rho^{\beta}
  \psi.
\label{eq: enhanced bilinears}
\end{eqnarray}

Since the bilinears (\ref{eq: current bilinears}) are conserved currents,
they maintain their engineering scaling dimensions
even in the presence of the gauge fluctuations.
On the other hand, the  bilinears
of the form (\ref{eq: enhanced bilinears})
(``enhanced bilinears'')
develop anomalous dimensions when we include
the gauge fluctuations.

Let us give some relevant examples of bilinears.
The $\sqrt{3} \times \sqrt{3}$ magnetically ordered state
(shown in Fig.~\ref{fig: kagome_lattice})
corresponds to the staggered charge density wave of vortices,
which is obtained by adding the corresponding staggered chemical
potential (i.e., $\pm \delta\mu$ on the up/down kagome triangles).
Using the listed symmetries, it is readily verified that
the following two bilinears in the expression
\begin{equation}
B_{\sqrt{3} \times \sqrt{3}} =
 \alpha\; \bar\psi \sigma^0 \mu^0 \tau^3 \rho^0 \psi
+ \beta\; \bar\psi \sigma^3 \mu^3 \tau^0 \rho^3 \psi
\end{equation}
transform in the manner expected of the staggered chemical potential
(the coefficients $\alpha$ and $\beta$ are generically independent).
This can be also checked explicitly by writing such chemical potential
in terms of the continuum fields obtaining some $t'$-dependent
coefficients $\alpha$ and $\beta$.
In the similar vein, the $q=0$ state
(also shown in Fig.~\ref{fig: kagome_lattice})
corresponds to the uniform
chemical potential for vortices on the up and down kagome triangles,
which is realized with the following bilinears
\begin{equation}
B_{q = 0} =
 \alpha'\; \bar\psi \sigma^0 \mu^3 \tau^3 \rho^3 \psi
+ \beta'\; \bar\psi \sigma^3 \mu^0 \tau^0 \rho^0 \psi ~.
\end{equation}
Note that each $B_{ \sqrt{3} \times \sqrt{3}}$ and $B_{q = 0}$
contains an enhanced bilinear of the QED$_3$ theory, and therefore
both orders are ``present'' in the AVL phase as enhanced critical
fluctuations.  It is in such sense that the enhanced bilinears
encode the potential nearby orders.

With the 
table~\ref{tab: Summary of symmetry transformations for cont fermions} 
at hand, it is a simple matter to check that
spatial symmetries
(translation, rotations, and reflection)
and particle-hole symmetry
prohibit all fermion bilinears
from appearing in the effective action
(\ref{eq: effective action}),
except $\bar{\psi}\psi$
and $\bar{\psi}\sigma^3 \mu^3 \tau^3 \rho^3 \psi$.
If we are allowed to require the invariance under
$\mathcal{T}_{\mathrm{ferm}}$, then these bilinears
would be prohibited as well.
However, even if we do not use $\mathcal{T}_{\mathrm{ferm}}$,
we exclude these bilinears from the continuum theory using
the following argument from Ref.~\onlinecite{Alicea05b}:
Consider first adding the bilinear $\bar{\psi}\psi$ to the
action and analyze the resulting phase for the original
spin model. 
This bilinear opens a gap in the fermion spectrum, and
proceeding as in Ref.~\onlinecite{Alicea05b} we conclude that
this phase 
is in fact a chiral spin liquid that breaks the physical time
reversal.
Therefore, if we are interested in a time-reversal invariant
spin liquid, we are to exclude this term.  The situation with
the $\bar{\psi}\sigma^3 \mu^3 \tau^3 \rho^3 \psi$ term is less clear 
since by itself it would lead to small Fermi pockets, 
and it is then difficult to deduce the physical state of the original
spin system.  In the presence of both terms, depending on 
their relative 
magnitude one may have either a gap or Fermi pockets.
However, we can plausibly argue that these pockets tend to be 
unstable towards a gapped phase that is continuously connected to the
same chiral phase obtained when the mass term $\bar{\psi}\psi$ dominates.
Since we are primarily interested in the states that are not chiral
spin liquid (e.g., states that appear from the AVL description by
spontaneously generating some other mass terms like 
$B_{\sqrt{3} \times \sqrt{3}}$ or $B_{q=0}$), 
we drop the bilinears 
$\bar{\psi}\psi$ and $\bar{\psi}\sigma^3 \mu^3 \tau^3 \rho^3 \psi$
from further considerations.
The validity of this assumption and the closely related issue of 
neglecting irrelevant higher-derivative Chern-Simons terms from the 
final AVL action are the main unresolved questions about the AVL 
approach (see Ref.~\onlinecite{Alicea05b} for some discussion).

As another example of the application of the derived PSG,
we write explicitly combinations of enhanced bilinears
that contribute to 
$S_j^z \sim (\nabla \times a)_j/(2\pi) + G_j + F_j + \dots$:
\begin{eqnarray}
G_j &=& g_j 
\left( e^{i {\bf Q} \cdot {\bf r}_j} 
      e^{-i {\pi}/{12}} B_{Q}^+
             + \Hc \right) ~,
\\
F_j &=& f^1_j \left(e^{i {\bf P}_1 \cdot {\bf r}_j} 
              e^{i {5\pi}/{12}}
               B_{P_1}^+ + \Hc \right)   
\nonumber \\
    && + f^2_j \left(e^{i {\bf P}_2 \cdot {\bf r}_j} 
               e^{i {5\pi}/{12}}
               B_{P_2}^+ + \Hc \right) 
\nonumber \\
    && + f^3_j \left(e^{i {\bf P}_3 \cdot {\bf r}_j} 
               e^{i{5\pi}/{12}}
               B_{P_3}^+ + \Hc \right) ~.
\end{eqnarray}
Here 
$j$ refers to the ``basis'' labels in the unit cell
consisting of four sites in the extended model shown
in Fig.\ \ref{fig: kagome_lattice};
the wave vectors are the ones shown in
Fig.~\ref{fig: enhbilin.eps},
while the corresponding bilinears are
$B_Q^+ =  \bar\psi \mu^2 \tau^+ \rho^0 \psi$,
$B_{P_1}^+ = \bar\psi \mu^3 \tau^+ \rho^2 \psi$,
$B_{P_2}^+ = \bar\psi \mu^2 \tau^+ \rho^3 \psi$,
$B_{P_3}^+ = \bar\psi \mu^0 \tau^+ \rho^2 \psi$.
The above is to be interpreted as an expansion of the microscopic
operators $S_j^z$ defined on the original spin lattice sites
in terms of the continuum fields in the theory.
It is straightforward but tedious to verify that one can choose real
parameters $g_j$ and $f^{1,2,3}_j$ so that $G$ and $F$ have identical
transformation properties with $\nabla \times a$ and therefore indeed
contribute to $S^z$. 
We thus conclude that $S^z$ has enhanced correlations at the
wavevectors $\pm{\bf Q}$, $\pm{\bf P}_{1,2,3}$ as claimed in the
main body.


\begin{thebibliography}{99}
\bibitem{remark specific heat}
When Dirac fermions and photons are decoupled, both of them
give rise to $T^2$ contributions to specific heat due to
their relativistic energy spectra. The effect of gauge
fluctuations
can be studied systematically for large numbers of fermion
flavor $N_f$.
(See also discussions in Ref.\ \onlinecite{Ran06}.)
One way to see how gauge interactions affect the specific
heat is to
integrate out fermions and derive the propagator for
dressed photons,
which becomes $\sim 1/\sqrt{|\omega|^2 + k_x^2 + k_y^2}$.
\cite{Rantner02, Franz03, Hermele05}
Calculating the free energy of such photons at finite
temperature,
one can then see that the specific heat still behaves as
$\sim T^2$,
although the "damped" dynamics of photons does modify the
proportionality constant.
\bibitem{remark 2}
For example, 
Dzyaloshinskii-Moriya interactions can provide the
desired easy
plane anisotropy. However, typically such interactions
also break some lattice
symmetries, such as inversions.
In terms of our effective field theory approach,
they could then modify the set of fermion bilinears
allowed by symmetry, which may lead to an instability of
the critical
spin liquid state we discuss.
Nevertheless, there may still be a range of energy scales
over which the
spin liquid might provide a reasonable description of an
experimental
system such as herbertsmithite.
These issues were addressed in Refs.\ 
\onlinecite{Alicea05a,Alicea05b} that discuss
geometrically frustrated quantum magnets on the triangular
lattice, motivated by Dzyaloshinskii-Moriya interaction in
CsCuCl.
\end{thebibliography}
\end{document}